\documentclass[twocolumn,10pt]{article}

\usepackage[margin=1in]{geometry}
\usepackage[T1]{fontenc}
\usepackage[utf8]{inputenc}

\usepackage{amsmath,amssymb}

\usepackage{graphicx}
\usepackage{booktabs}
\usepackage{multirow}
\usepackage{float}
\usepackage{subcaption}   
\usepackage{mdframed}

\usepackage{multicol}

\usepackage[table,xcdraw]{xcolor}

\usepackage{arydshln}

\usepackage{listings}

\usepackage[hidelinks]{hyperref}

\lstdefinelanguage{Prompt}{
  morekeywords={Output, JSON, label, return},
  sensitive=false,
  morecomment=[l]{//},
  morestring=[b]",
}

\title{Leveraging Language Models to Discover Evidence-Based Actions for OSS Sustainability}

\author{
Nafiz Imtiaz Khan\thanks{Corresponding author: \texttt{nikhan@ucdavis.edu}}, Vladimir Filkov \\
Department of Computer Science, University of California, Davis, CA 95616, USA
}

\date{} 

\begin{document}

\maketitle

\begin{abstract}
When successful, Open Source Software (OSS) projects create enormous value, but most never reach a sustainable state. Recent work has produced accurate models that forecast OSS sustainability, yet these models rarely tell maintainers what to do: their features are often high-level socio-technical signals that are not directly actionable. Decades of empirical software engineering research have accumulated a large but underused body of evidence on concrete practices that improve project health.

We close this gap by using LLMs as evidence miners over the SE literature. We design a RAG-pipeline and a two-layer prompting strategy that extract researched actionables (ReACTs): concise, evidence-linked recommendations mapping to specific OSS practices. In the first layer, we systematically explore open LLMs and prompting techniques, selecting the best-performing combination to derive candidate ReACTs from 829 ICSE and FSE papers. In the second layer, we apply follow-up prompting to filter hallucinations, extract impact and evidence, and assess soundness and precision.

Our pipeline yields 1,922 ReACTs, of which 1,312 pass strict quality criteria and are organized into practice-oriented categories connectable to project signals from tools like APEX. The result is a reproducible, scalable approach turning scattered research findings into structured, evidence-based actions guiding OSS projects toward sustainability.
\end{abstract}

\textbf{Keywords:} Research Actionable, Open Source Software, Sustainability, Literature Review

\section{Introduction}

Open Source Software (OSS) underpins a multi-billion-dollar global ecosystem and is a foundational component of modern software systems. Most large technology companies rely extensively on OSS in their production infrastructure. A recent Harvard Business School report estimates that recreating the OSS components currently embedded in commercial software would cost approximately \$8.8 trillion \cite{14_layne2024}. This figure represents demand-side value, while supply-side estimates suggest that OSS development costs are on the order of \$4.2 billion, despite OSS being used by over 96\% of commercial software products. Complementing this view, another industry report projects that the global OSS market will reach \$50 billion by 2026 \cite{15_gitnux2024}.

Despite its economic and societal importance, OSS faces a persistent sustainability challenge. Prior studies report that over 90\% of OSS projects are eventually discontinued, with failure rates particularly high among smaller and younger projects \cite{13_coelho2017modern,41_schweik2012internet}. This challenge motivates two long-standing and interrelated research problems: (a) \emph{predicting} whether an OSS project is likely to remain sustainable, and (b) \emph{identifying concrete actions} that project maintainers can take to intervene when sustainability risks emerge.


\textbf{The Prediction-Action Gap.} A substantial body of prior work has focused on the first problem, predicting OSS sustainability, using statistical and machine learning models over socio-technical features. Ghapanchi et al.\ \cite{17_ghapanchi2015predicting} analyzed longitudinal data from 1{,}409 OSS projects and proposed predictive models based on factors such as developer activity, release frequency, project age, and collaboration patterns. Yin et al.\ \cite{16_yin2021sustainability} introduced an LSTM-based deep learning model that achieved over 93\% accuracy using 18 socio-technical features. More recently, Xiao et al.\ \cite{18_xiao2023early} employed 64 early-stage variables to forecast OSS sustainability during project inception.

While these approaches demonstrate strong predictive performance, they provide limited guidance on the second problem: \emph{how} practitioners should respond to predicted risks. The features that drive these predictions, such as call-graph complexity, aggregate code churn, or developer network centrality, are often abstract or indirectly measurable signals of project health. They tell maintainers \emph{what} might be going wrong, but not \emph{what to do} about it. As a result, there remains a critical gap between sustainability prediction and actionable intervention: even when a model forecasts that a project is at risk, maintainers are left without clear, evidence-based guidance on which practices to adopt or change.

\textbf{From Research Findings to Actionable Recommendations.} In this work, we address this prediction-action gap by extracting \emph{evidence-based actionable recommendations} from the OSS research literature. We leverage the concept of \textbf{ReACTs} (Researched Actionables), introduced by Khan and Filkov~\cite{19_khan2024models}: concise, practice-oriented recommendations grounded in empirical findings from peer-reviewed software engineering studies. Rather than predicting sustainability outcomes directly, ReACTs operationalize research insights into concrete actions that maintainers can plausibly undertake, such as improving onboarding documentation, adjusting contribution workflows, restructuring release practices, or adopting specific code review techniques. Each ReACT is explicitly linked to its source study, the empirical evidence supporting it, and the stated impact on overall project health.

Recent advances in LLMs provide a new opportunity to scale the extraction of such actionables from the vast corpus of software engineering research. LLMs have demonstrated strong capabilities in summarization, reasoning, and structured information extraction \cite{20_llmcapabilities}, and prompt engineering techniques can further guide models toward domain-specific outputs \cite{21_promptbasic}. However, naively applying LLMs to scientific text risks generating vague, unsupported, or non-actionable suggestions, a problem we address through structured validation.

To mitigate these risks, we propose a structured LLM-based pipeline for extracting high-quality ReACTs from OSS research articles. Our approach combines Retrieval-Augmented Generation (RAG) with a two-layer refinement and reliability-checking process designed to improve factual grounding, clarity, and actionability. In the first layer, we systematically evaluate multiple open-source LLMs and prompting techniques to identify the best-performing combination for ReACT extraction. In the second layer, we apply follow-up prompting to filter hallucinations, extract explicit statements of impact and empirical evidence, and assess each ReACT for soundness (logical consistency) and precision (clarity and specificity). We apply this pipeline at scale to 829 OSS papers from top SE conference venues (ICSE and FSE), yielding a curated catalog of ReACTs organized into practice-oriented categories that can be mapped to project health signals via the APEX sustainability framework.
\newline
\newline
\textbf{Contributions and Relation to Prior Work.}
This study builds on and substantially extends prior work presented in a FSE-IVR idea paper \cite{19_khan2024models}. That earlier work manually derived 105 actionables from a fixed set of 186 papers and mapped them to 18 preselected socio-technical features from \cite{16_yin2021sustainability}. The present work makes four key advances over that preliminary study:

\begin{enumerate}
    \item \textbf{Corpus-scale LLM-based extraction:} We scale ReACT extraction from 186 manually-reviewed papers to 829 ICSE/FSE articles using LLMs with a RAG pipeline, systematically evaluating four open-source models (Llama3-8B, Mixtral-8x7B, MistralLite-7B, Mistral-Nemo-12B) and three prompting techniques (Zero-Shot, Chain-of-Thought, Reason+Action) to identify the best-performing combination.
    
    \item \textbf{Two-layer refinement and reliability validation:} We introduce an automated pipeline that applies follow-up prompting to filter hallucinations (ReACTs not actually mentioned in source papers), extract explicit impact and evidence statements, and evaluate each ReACT for soundness and precision, yielding 1,922 validated ReACTs, of which 1,312 meet strict quality criteria.
    
    \item \textbf{Category-level organization:} Rather than mapping to fixed predictive features, we organize ReACTs into eight practice-oriented categories (e.g., New Contributor Onboarding, Code Standards and Maintainability, Community Collaboration) that directly correspond to maintainer actions and can be flexibly connected to project states.
    
    \item \textbf{Operational integration with ASFI:} We demonstrate practical applicability through detailed case studies of Apache Software Foundation Incubator (ASFI) projects, showing how ReACTs can be selected based on APEX-identified sustainability turning points to provide targeted, evidence-based guidance during critical project phases.
\end{enumerate}

Together, these contributions move beyond manual, feature-bound actionables toward a scalable, reproducible, and actionable bridge between OSS research and practice. The result is a structured catalog of evidence-based actions that practitioners can use to guide OSS projects toward sustainability, grounded in decades of empirical software engineering research and validated through rigorous LLM-based extraction and refinement.

\section{Background}

\subsubsection{Language Models for Software Engineering}
With the emergence of open-source LLMs, SE researchers and developers are utilizing LLMs for various SE tasks. According to a survey study by Zheng et al.~\cite{2_zheng2023towards}, practitioners are using LLMs for seven different SE tasks: a) code generation, b) code summarization (i.e., generating comments for code), c) code translation (i.e., converting code from one programming language to another), d) vulnerability detection (i.e., identifying and fixing defects in programs), e) code evaluation and testing, f) code management (i.e., code maintenance activities such as version control), and g) Q\&A interaction (i.e., using Q\&A platforms such as StackOverflow). Rasnayaka et al.~\cite{5_rasnayaka2024empirical} found that 40.5\% of the total student teams are using LLMs in academic projects, and students with higher coding skills are more inclined to use LLMs. Hou et al.~\cite{4_hou2023large} found that there are traces of using LLMs on 85 different SE tasks, among them Code Generation and Program Repair are the most prevalent tasks for employing LLMs in software development and maintenance activities. They also found that practitioners are using different SE datasets (data on source code, bugs, patches, code changes, test suites, Stack Overflow, API documentation, code comments, and project issues) for sophisticated prompting/fine-tuning the models. Recent work by Babar et al.~\cite{babar2026opensource} demonstrated that open-source LLMs, when enhanced with RAG and fine-tuning, can effectively answer technical queries from StackExchange platforms, offering scalable and cost-effective alternatives to proprietary models for developer assistance tasks.

The above studies point to the LLMs' effectiveness in handling a wide variety of software engineering tasks and data.

\subsubsection{Language Models for Evidence Extraction}
LLMs are being employed in various domains to extract evidence from documents and scientific literature. For instance, Gartlehner et al.~\cite{6_gartlehner2024data} utilized the web version of Claude 2 to extract 160 data elements from 10 open-access Randomized Control Trial (RCT). The study found that Claude 2 can extract data with an overall accuracy of 96.3\% (6 errors out of 160 data elements). Likewise, the study conducted by Huang et al. \cite{9_huang2024critical} extracted structured information from over 1000 lung cancer and 191 pediatric osteosarcoma pathology reports. The authors utilized ChatGPT-3.5 (gpt-3.5-turbo-16k) through OpenAI's API for batch querying. The evaluation showed ChatGPT-3.5 achieved 89\% overall accuracy for lung cancer classifications, outperforming traditional NLP methods, and 98.6\% and 100\% accuracy for osteosarcoma grades and margin status, respectively. 

In another study, the authors proposed and evaluated a zero-shot strategy using LLMs to retrieve and summarize relevant unstructured evidence from patient Electronic Health Records (EHRs) based on specific queries~\cite{10_ahsan2023retrieving}. They compared their LLM-based approach against a pre-LLM information retrieval baseline, with expert evaluation showing a consistent preference for the LLM outputs. In another separate study, Patiny and Godlin~\cite{7_patiny2023automatic} outlined a method for LLM-based automatic extraction of experimental FAIR (Findable, Accessible, Interpretable, Reusable) data of molecules from literature published in the domain of chemistry. Here also, authors used ChatGPT 3.5 turbo API version to conduct the experiments and could extract 74\% of the data from published papers. Similarly, in the field of mental health, Alhamed et al.~\cite{8_alhamed2024using} used LLM to extract justification of a pre-assigned gold label from a suicidality dataset containing Reddit posts labeled with the suicide risk level. The authors used \textit{Llama 7b} for experimenting and achieved a precision of 0.96. 

According to previous studies, LLMs are widely used in various fields to extract evidence from reports and published literature. However, no such research has been conducted in SE to extract evidence or information from SE literature or technical reports, highlighting a plausibility and urgency for exploration.

\section{Research Questions}
LLMs have shown potential in various tasks, but their ability to extract actionable insights from the scientific literature remains under-explored. Given their advanced capabilities, we posit that LLMs can effectively generate actionable recommendations that are both relevant and useful for enhancing practices. Understanding their efficacy in this domain can unlock new avenues for automating knowledge extraction from research articles.

\vspace{0.25cm}
\begin{mdframed}[linewidth=0.5pt, linecolor=black, backgroundcolor=blue!10]
$\textbf{RQ}_{\textcolor{red}{1}}$ Can we use LLMs to effectively derive evidence-based actionable recommendations from scientific literature? 
\end{mdframed} 
\vspace{0.25cm}

When LLMs generate actionable recommendations, the quantity and quality of these outputs are crucial. Poor-quality actionables can mislead both practitioners and researchers, underscoring the importance of evaluating whether LLM-derived recommendations meet the necessary standards for practical use in enhancing OSS sustainability.

\vspace{0.25cm}
\begin{mdframed}[linewidth=0.5pt, linecolor=black, backgroundcolor=blue!10]
\textbf{RQ$_2$:} How many evidence-based actionable recommendations can LLMs derive from scientific literature published in top SE venues (ICSE and FSE) in the domain of OSS, and what is the reliability of those derived actionables?
\end{mdframed}
\vspace{0.25cm}

Even if LLMs can generate high-quality actionables, their value depends on their implementation in real-world settings. The gap between theoretical recommendations and their practical application often remains unaddressed, and understanding how these actionables can be effectively integrated into software engineering practices, e.g, through existing traces or workflow monitoring tools, is vital for achieving tangible improvements in OSS projects. 

\vspace{0.25cm}
\begin{mdframed}[linewidth=0.5pt, linecolor=black, backgroundcolor=blue!10]
$\textbf{RQ}_{\textcolor{red}{3}}$ How can the actionables be presented meaningfully in practice?
\end{mdframed} 
\vspace{0.25cm}

\section{Methodology}
Here, we describe a number of diverse methods we utilize to answer our research questions. The overall study methodology is presented through a conceptual diagram in Fig.  \ref{fig:ConceptualDiagram}. 

\subsection{Selection of Relevant Articles}
Most SE researchers present their best work at prestigious conferences. Thus, our focus is on publications in top-tier SE conference venues. Specifically, ICSE (International Conference on Software Engineering) and ESEC/FSE (ACM Joint European Software Engineering Conference and Symposium on the Foundations of Software Engineering), which consistently rank highest in the software engineering domain~\cite{28_conferenceranks, 29_core}. Our study utilized articles presented at ICSE and FSE from the 21\textsuperscript{st} century (from June 2000 to February 2025).

In the given range, we found a total of 7593 published articles in ICSE and FSE. Given our interest in deriving actionable insights to enhance the sustainability of OSS projects, we further refined our selection criteria. We included only those articles whose titles or abstracts contained the strings ``Open Source'' or ``OSS,'' which resulted in 569 ICSE articles and 260 FSE. Thus, our dataset comprises those 829 conference articles.

\subsection{ReACT Definition and Examples}
\label{sec:react_definition}

Before describing our extraction methodology, we first clarify what constitutes a ReACT (Researched Actionable). A \textbf{ReACT} is a concise, evidence-based recommendation extracted from peer-reviewed software engineering research that specifies a concrete action maintainers can take to improve OSS project health. Each ReACT consists of three components:

\begin{enumerate}
    \item \textbf{Action:} A clear, implementable practice (e.g., ``Implement automated code linting in the CI pipeline'')
    \item \textbf{Impact:} The stated effect on project outcomes (e.g., ``Reduces code review time and improves code quality'')
    \item \textbf{Evidence:} The empirical basis from the source study (e.g., ``Empirical analysis of OSS projects showed reduction in defect density'')
\end{enumerate}

\noindent\textbf{Example ReACT 1 (Complete: Both Action \& Impact Present):}
\begin{itemize}
    \item \textit{Action:} ``Encourage mentors to provide feedback on pull requests (PRs) by acknowledging their efforts and contributions''
    \item \textit{Impact:} ``Increases newcomer retention and accelerates their path to becoming regular contributors''
    \item \textit{Evidence:} ``Mixed-methods study of 12 Apache projects found that acknowledged newcomers were 2.3× more likely to make subsequent contributions''
    \item \textit{Source:} Feng et al.~\cite{81_feng2022case}
\end{itemize}

\noindent\textbf{Example ReACT 2 (Action-only, no explicit impact):}
\begin{itemize}
    \item \textit{Action:} ``Consider synchronizing branches before applying refactorings that are likely to be incompatible with changes made in parallel''
    \item \textit{Impact:} (Not explicitly stated in source article)
    \item \textit{Evidence:} ``Analysis of 10,000+ merge scenarios in 30 projects showed that pre-synchronization reduced merge conflicts by 41\%''
    \item \textit{Source:} Oliveira et al.~\cite{82_oliveira2023code}
\end{itemize}

These examples illustrate that not all extracted ReACTs are equally complete or actionable, motivating our two-layer refinement and reliability assessment process (described below).

\subsection{LLMs Task Definition}
To address the RQs, we defined three specific tasks while adopting an experimental consideration, which are discussed as follows:

\begin{figure*}[]
\includegraphics[width = \linewidth, height=10.4cm]{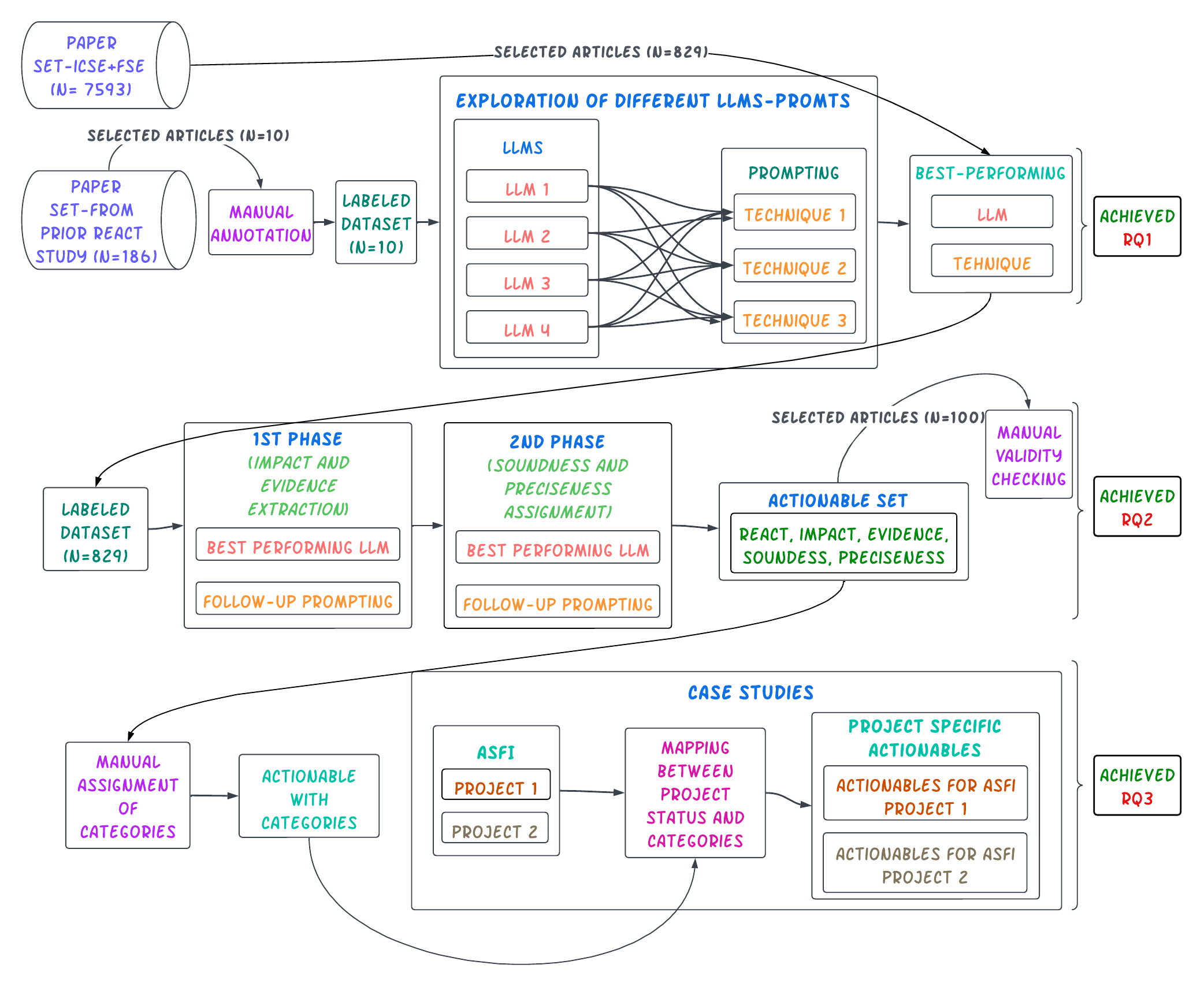}
\caption{Conceptual Diagram Illustrating the Methodology}
\label{fig:ConceptualDiagram}
\end{figure*}

\subsubsection{ReACT Derivation} 
In this task, LLMs were provided with a prompt and the full text of an article, along with instructions for extracting ReACTs. The ReACTs are the basic units of our analyses, inspired by a prior limited study~\cite{19_khan2024models}. That study analyzed 186 articles and identified ReACTs from 27 of them. From the study's provided list of 186 papers, we randomly selected a small subset (n=10) for manual annotation. Two researchers collaborated to annotate these articles for ReACTs, successfully identifying ReACTs in six of them, yielding 42 candidate ReACTs for evaluation. The list of the ten articles, along with their annotations, can be found in Section 4 of the Appendix.

Using the hand annotations as a reference, we evaluated different combinations of LLMs and prompting techniques (detailed below) to determine the most effective model and prompting strategy for the ReACT derivation task. The top-performing LLM and prompting technique was then applied to the full dataset of 829 articles of this study to derive all ReACTs.

\subsubsection{ReACT Refinement} 
To improve the derived ReACTs, we implemented a refinement layer aimed at identifying relevant information about them. This layer extracts ``impact'' and ``evidence'' for each ReACT and ensures that the actionables are backed by sufficient contextual information. It also filters hallucinated ReACTs, those generated by the LLM but not actually present in the source article.

\subsubsection{ReACT Reliability Assessment} 
This task involved assessing the reliability of the finalized, refined ReACTs using predefined criteria. The purpose of this task is to ensure whether the generated ReACTs are logically consistent (soundness) and directly adaptable (preciseness).

\subsubsection{Experimental Consideration}
In our study, we performed all experiments while keeping a constant model temperature (\textit{temp} = 0). Temperature in the context of LLMs is a hyperparameter that controls the randomness of the model's output~\cite{51_temperature}. A lower temperature (closer to 0) makes the model's responses more deterministic and focused, while a higher temperature increases randomness and creativity. The other hyperparameters of the model, such as: \textit{top\_p, top\_k, min\_p, and repeat\_penalty} were kept at default values.

\subsection{Evaluation Metrics}
Due to the lack of standardized evaluation measures for LLMs, we use a variety of metrics to evaluate both the LLM performance and the reliability of the predictions.

\subsubsection{LLM Performance: Metrics for Model Selection}
\label{sec:llm_metrics}

For selecting the best-performing LLM and prompting technique on our small labeled set, we employed four quantitative evaluation metrics inspired by prior literature~\cite{46_ahmed2024studying, 46_virk2024enhancing}. The selected metrics can evaluate the LLM-generated response while giving gold annotation as a reference. Among the selected metrics, two are text-similarity-based (\textit{BLEU-4} and \textit{ROUGE-L}), and the other two are semantic-based (\textit{BERTScore} and \textit{METEOR}). Semantic-based metrics evaluate text by analyzing meaning and context, considering factors like coherence and relevance~\cite{47_semantic}. In contrast, similarity-based metrics primarily focus on lexical overlap or statistical similarity between texts~\cite{47_similarity}.

These metrics are computed against a \emph{single gold phrasing} per ReACT from our manual annotations. Because valid ReACTs can be expressed in many paraphrased forms, BLEU, ROUGE, BERTScore, and METEOR scores provide only a weak proxy for true ReACT quality. We use these metrics solely for relative model selection, to identify which LLM and prompting combination performs best on our small labeled set. The \emph{primary evidence of quality} comes from our subsequent human sanity checks and inter-annotator agreement (IAA) scores on a stratified sample of 100 ReACTs (described in Section~\ref{sec:sanity_check}).

Although there is no universally accepted fixed threshold for these metrics, recent studies from top-ranked venues suggest that a \textit{BERTScore} above 0.6, a \textit{BLEU-4} score above 0.4, a \textit{ROUGE-L} score above 0.4, and a \textit{METEOR} score above 0.5 can be considered indicative of very good performance~\cite{46_ahmed2024studying, 46_ahmed2024automatic, 46_al2023extending, 46_virk2024enhancing}.

\textbf{\textit{BLEU-4.}} \textit{BLEU-4} (Bilingual Evaluation Understudy) calculates the geometric mean of n-gram precisions up to 4-grams~\cite{44_bleu}. This metric captures local fluency and adequacy by measuring the overlap of short word sequences (up to 4 words long) between the generated text and reference translations. 

\textbf{\textit{ROUGE-L.}} \textit{ROUGE-L} (Recall-Oriented Understudy for Gisting Evaluation - Longest Common Subsequence) measures the longest common subsequence between the generated output and reference text~\cite{44_rouge}. Unlike n-gram-based metrics, ROUGE-L captures sentence-level structure and allows for word order flexibility. 

\textbf{\textit{BERTScore.}} \textit{BERTScore} leverages pre-trained BERT embeddings to assess the semantic similarity between generated text and reference text~\cite{44_bert}. In our study, we utilized the BERT-base-uncased~\cite{58_BERTBaseUncased} variant of BERT, which has 12 layers and 768 hidden units; ``uncased'' means that the model does not differentiate between uppercase and lowercase letters.

\textbf{\textit{METEOR.}} \textit{METEOR} (Metric for Evaluation of Translation with Explicit ORdering) incorporates linguistic features to capture semantic equivalence and structural adequacy~\cite{44_meteor}. Its flexible matching strategy accounts for exact matches, stemming, synonymy, and paraphrasing, which provides a more nuanced evaluation of semantic similarity. 

\subsubsection{ReACT Reliability Assessment}
To assess the final set of ReACTs, we measure their reliability using two metrics: Soundness and Preciseness. 

\textbf{{Soundness.}}  
A ReACT is sound if it makes logical sense and has no contradictions. This means that all parts of the recommendation work together consistently. For instance, \textit{``Project Managers should use peer reviews and automated tools for code review''} is considered a sound ReACT as peer reviews and tools work together to improve code quality. On the other hand, \textit{``To improve user satisfaction, add new features without onboarding newcomers''}- is considered an unsound ReACT because adding features without testing can hurt user satisfaction, and not onboarding newcomers can reduce support quality.

\textbf{{Preciseness.}}  
A ReACT is precise if it is clear and easy to follow. It should be specific and leave no room for confusion. For example, \textit{``To attract newcomers, help them make their first contribution''} is considered precise as it gives a clear action to take. On the contrary, \textit{``To attract core developers, ensure high code quality''}- is considered imprecise as it does not explain how to ensure high code quality.

\subsection{LLM Selection}

\textit{Hugging Face} is one of the best platforms for hosting LLMs~\cite{71_jain2022hugging}. However, it is quite challenging to identify the most suitable set of models for a specific task, given the vast number of models available on \textit{Hugging Face}. Thus, to address this challenge, we considered several factors.

\subsubsection{Selection Criteria}

\textbf{\textit{Privacy.}} As maintaining the privacy of the papers is of paramount importance in our study, we selected LLMs that can be run locally without internet reliance. We identified several open-source LLMs from renowned tech companies like Meta~\cite{meta_metaverse_2023}, Google~\cite{google_2024}, Microsoft~\cite{microsoft_2024}, and xAI~\cite{xai_2024}, which can operate on personal computers or local servers without cloud services.

\textbf{\textit{Model Types.}} There exist different types of the same model, such as \textit{instruct}~\cite{35_instruct}, \textit{conversational}~\cite{33_conersational}, and \textit{completion models}~\cite{34_completion}. However, in this study, we selected \textit{instruct} type models, as these models are particularly optimized for following detailed prompts~\cite{35_instruct}.

\textbf{\textit{Context Window.}} A critical factor in selecting LLMs is the context window, which represents the maximum number of tokens the model can process at once~\cite{30_context_window}. We calculated the token size for each article; the maximum token size is 24k. To ensure consistency in the ReACT derivation task, we selected LLMs with context windows exceeding (24 + 4)k tokens (here, we reserved 4k tokens for the prompt). Open-source LLMs have context windows ranging from 2k to 128k tokens, with intermediate values at 16k, 32k, and 64k. Thus, we considered models having a context window of at least 32k. 

\textbf{\textit{Hardware Constraints}} The models we can eventually run depend significantly on the availability of computing resources. We conducted all the experiments of our study on a local server, which has a specific configuration (detailed in section C of the Appendix). Therefore, we excluded models that can not be run on our local server (requiring more than 12GB VRAM).

Taking all factors into account (privacy, model type, hardware compatibility, and context window), we identified four models as potential candidates: \textit{Llama3-8B, Mixtral-8x7B, MistralLite-7B, and Mistral-Nemo-12B}.

\subsubsection{Selected Models}

\textbf{{\textit{Llama3-8B}.}} Developed by Meta~\cite{63_MetaAbout}, \textit{Llama3 8B} is a language model with 8 billion parameters and a default context size of 8k tokens~\cite{27_llama}. We utilized the version of the model which has a context window of 32k tokens~\cite{64_MetaLlama38B} and is contributed by Nurture AI~\cite{65_NurtureAI}.

\textbf{{\textit{Mixtral-8x7B}.}} The model \textit{Mixtral 8x7B} was developed by Mistral AI~\cite{59_MistralAI} with 46.7 billion parameters and a default context window of 32k tokens~\cite{25_mixtral}. Notably, it uses a mixture-of-experts architecture, which allows for efficient scaling and potentially better performance than similarly-sized models.

\textbf{{\textit{MistralLite-7B}.}} The model \textit{MistralLite} is contributed by Amazon AWS~\cite{60_AWS}, which has 7 billion parameters and a context window of 32k~\cite{61_MistralLite}. This model is based on the original Mistral 7B model~\cite{62_Mistral7BDiscussion} (created by Mistral AI~\cite{59_MistralAI}). By utilizing adapted rotary embedding and sliding window during training, the model MistralLite performs significantly better on several long context tasks while keeping the same model structure as the original model.

\textbf{{\textit{Mistral-Nemo-12B}.}} The model \textit{Mistral-Nemo 12B} was developed by Mistral AI~\cite{59_MistralAI}, which features 12 billion parameters and a default context window of 128k tokens~\cite{69_mistralnemo}. Its large parameter count and substantial context window contribute to its robust performance across diverse NLP benchmarks~\cite{69_mistralnemobenchmarks}.

\subsection{Prompt Engineering}

A prompt in LLM is the input text or instruction given to the model to elicit a specific response or generate relevant output~\cite{93_cheng2024novel}. In this study, we adapted several well-defined prompting techniques and performed iterative prompt engineering, which is the process of designing, refining, and optimizing input prompts to effectively guide LLMs in generating desired outputs~\cite{37_marvin2023prompt}. For designing the prompts, we followed the 26 principled instructions provided by Bsharat et al.~\cite{38_prompt_guidelines}, which are effective in getting the desired response by LLMs. Also, we appended an emotional stimulus with the prompts (wherever applicable), as they can significantly boost the performance of LLMs on generative tasks~\cite{39_emotional_stimuli}. 

We have performed prompt engineering for each of our defined tasks, which are described as follows:

\subsubsection{ReACT Derivation}

For this task, inspired by prior studies~\cite{99_schulhoff2024prompt, 100_gaur2024brief}, we adapted three prompting techniques: \textit{Zero-Shot, Chain-of-Thought, and Reason+Action}, which are discussed as follows.

\textbf{{\textit{Zero-Shot.}}} \textit{Zero-Shot} prompting allows LLMs to leverage general knowledge and tackle new problems based solely on task descriptions~\cite{40_zero}. The \textit{Zero-Shot} prompt used in our study contains 76 tokens, and the structure of the prompt is provided in Section 1.1 of the Appendix.

\textbf{{\textit{Chain-of-Thought.}}} Proposed by Wei et al.~\cite{40_CoT}, the \textit{Chain of Thought (CoT)} prompting technique guides the LLM model to break down complex problems into intermediate steps, which is similar to human-like reasoning processes. This approach involves prompting the model to articulate its thought process, step-by-step, before arriving at a final answer. The adapted \textit{CoT} prompt used in our study contains 336 tokens, and the structure of the prompt is provided in Section 1.2 of the Appendix.

Here, we divided the whole task into four steps: first, we instructed the LLMs to read each sentence of the provided article to look for imperative sentences or phrases that give commands, make requests, or offer instructions to direct or persuade someone to perform a specific action. Second, we instructed the LLMs to generate a list of actionables in the form of concise, clear, and unambiguous statements. Next, during our experiments, it was observed that LLMs sometimes generate slightly modified versions of the same actionable multiple times. Thus, we included step 3, which instructs the LLMs to review the list of recommendations and remove any duplicate items from the list. Finally, we introduced another step, where we instructed the LLMs to provide a confidence score (between 0 and 1), asking how confident they are that the provided actionable can help OSS projects become sustainable. We also asked the models to provide a brief explanation regarding the confidence score. The idea of asking the LLMs to generate a confidence score is a form of self-evaluation. According to prior literature~\cite{42_self_evaluation1, 42_self_evaluation2}, it can help the models overcome hallucinations and produce reliable responses. Likewise, the idea of ``self-confidence'' was adapted to our study to ensure LLMs do not generate spurious actionables. However, we did not include any example set of actionables in the given prompt to ensure that the model does not get biased by seeing examples. 

\textbf{{\textit{Reason+Action.}}} \textit{Reason+Action} is a prompting method that combines the strengths of \textit{CoT} prompting with actionable outputs~\cite{40_react}. This technique encourages the LLMs to first reason through a problem or situation, and then provide specific, implementable actions based on that reasoning. Unlike \textit{CoT}, this prompting forces the models to take some intermediate actions. Following the base structure of the \textit{CoT} prompt, the \textit{Reason+Action} prompt was set up. The adapted prompt contains 592 tokens, and the structure of the prompt is provided in Section 1.3 of the Appendix.

The \textit{Reason+Action} prompting approach employed in this study follows a structure similar to \textit{CoT} prompting. The prompt is organized into a series of thought, action, and observation cycles, which guide the models through a systematic process of information extraction and refinement. Like \textit{CoT}, initially the task is explicitly defined. This is followed by four distinct stages: \textit{Extraction}, where the model is instructed to carefully read the article and compile a list of actionable recommendations; \textit{Refinement}, where the initial list is reviewed and refined to ensure clarity; \textit{Duplicate Removal}, where the list is reviewed to see if there exist any duplicates (if found, the models are instructed to remove the duplicates); and \textit{Confidence Scoring}, where each recommendation is assigned a confidence score on a scale of 0 to 1, accompanied by a brief explanation.

\subsubsection{ReACT Refinement}
\label{sec:refinement_prompts}

ReACTs are evidence-based and should provide a tangible action when implemented. While our initial phase focused on deriving ReACTs, it did not explicitly extract ``impact'' and ``evidence'' supporting each actionable. This phase of the study performs this extraction and ensures that each action is accompanied by both impact and supporting evidence.

As the ReACT refinement task requires processing each actionable, we did not follow any traditional prompting technique here. Rather, to do the extraction, we implemented a follow-up prompting technique inspired by the methodology described by Polak and Morgan~\cite{polak2024extracting}. In their study, the authors utilized follow-up prompting and achieved significant accuracy in extracting materials data from scientific papers. We adapted this effective technique to our context, which involves a series of binary and descriptive questions.

The follow-up prompting technique is illustrated in Fig.~\ref{refinement}. Here, for this work, we employed the winner LLM from the ReACT derivation task. This technique was applied to each actionable derived from the articles. Initially, the content of the article was provided to the model, followed by a series of questions. 

\textbf{Explicit Prompt Sequence:}
\begin{enumerate}
    \item \textbf{Hallucination Check:} ``Answer YES or NO only. Does the article explicitly mention the following actionable: [ReACT text]?''
    \begin{itemize}
        \item If NO $\rightarrow$ discard ReACT (classified as hallucinated)
        \item If YES $\rightarrow$ proceed to impact extraction
    \end{itemize}
    
    \item \textbf{Impact Extraction:} ``Answer YES or NO only. Does the article explicitly state the impact(s) on open-source projects if this actionable is adopted?''
    \begin{itemize}
        \item If YES $\rightarrow$ ``What impact(s) does the article indicate would result from adopting this actionable? Provide the exact statement.''
        \item If NO $\rightarrow$ classify as ``actionable without impact''
    \end{itemize}
    
    \item \textbf{Evidence Extraction:} ``Answer YES or NO only. Does the article provide any empirical evidence to support its claims regarding the stated impact(s)?''
    \begin{itemize}
        \item If YES $\rightarrow$ ``Describe the empirical evidence presented in the article that supports the claim that the recommended action will produce the stated impact(s).''
        \item If NO $\rightarrow$ classify as ``actionable without evidence''
    \end{itemize}
\end{enumerate}


This ReACT refinement process, utilizing a sequence of follow-up prompts, effectively eliminates hallucinated ReACTs and ensures that the actionables are substantiated with both impact and evidence (if available).

\subsubsection{ReACT Reliability Assessment}
\label{sec:reliability_prompts}

\begin{figure*}
    \centering
    \begin{minipage}[]{0.55\textwidth}
        \centering
        \includegraphics[width=\textwidth, height=9cm]{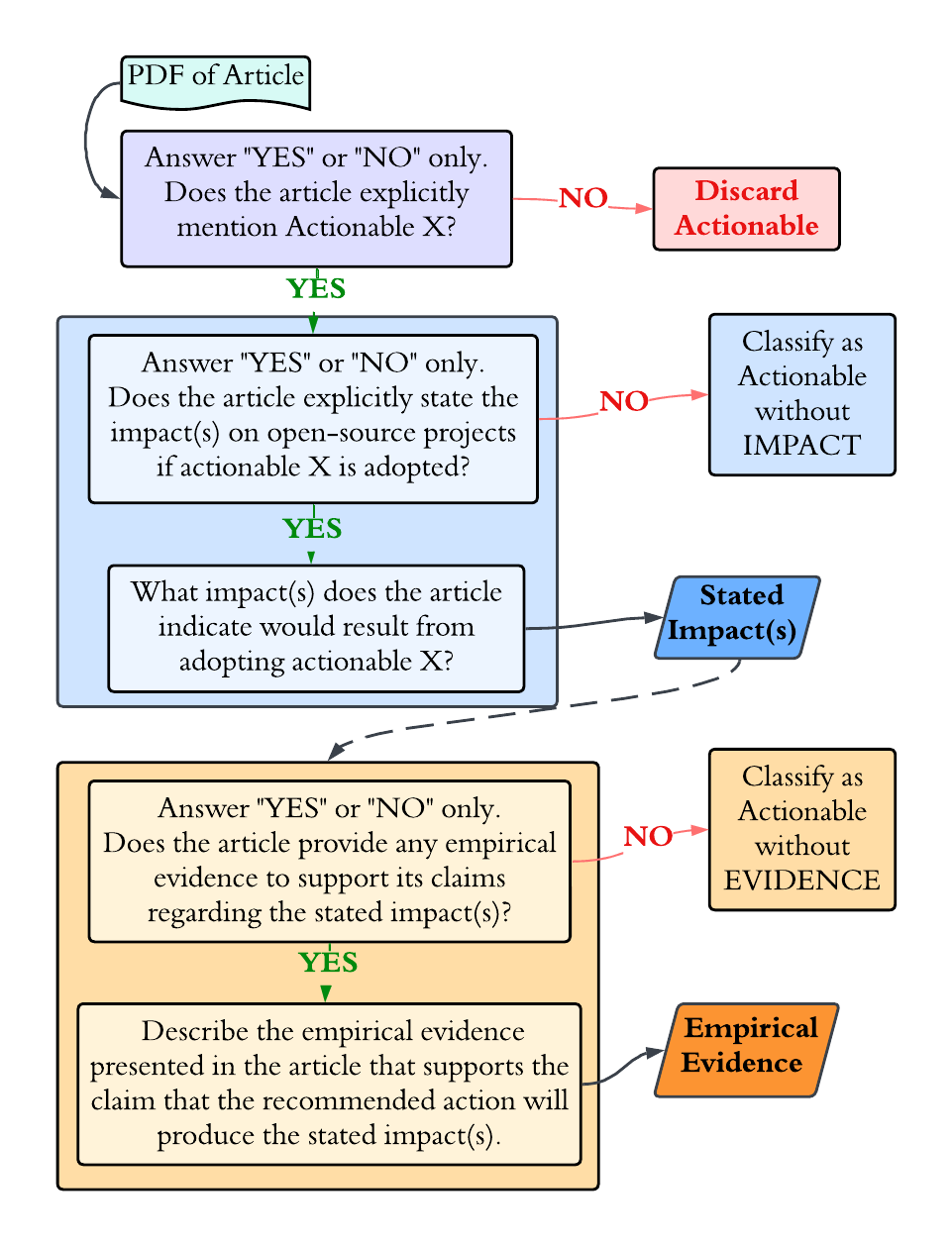}
        \caption{Flowchart of the Follow-Up Prompting Technique for ReACT Refinement}
        \label{refinement}
    \end{minipage}\hfill
    \begin{minipage}[]{0.43\textwidth}
        \centering
        \includegraphics[width=\textwidth, height=9cm]{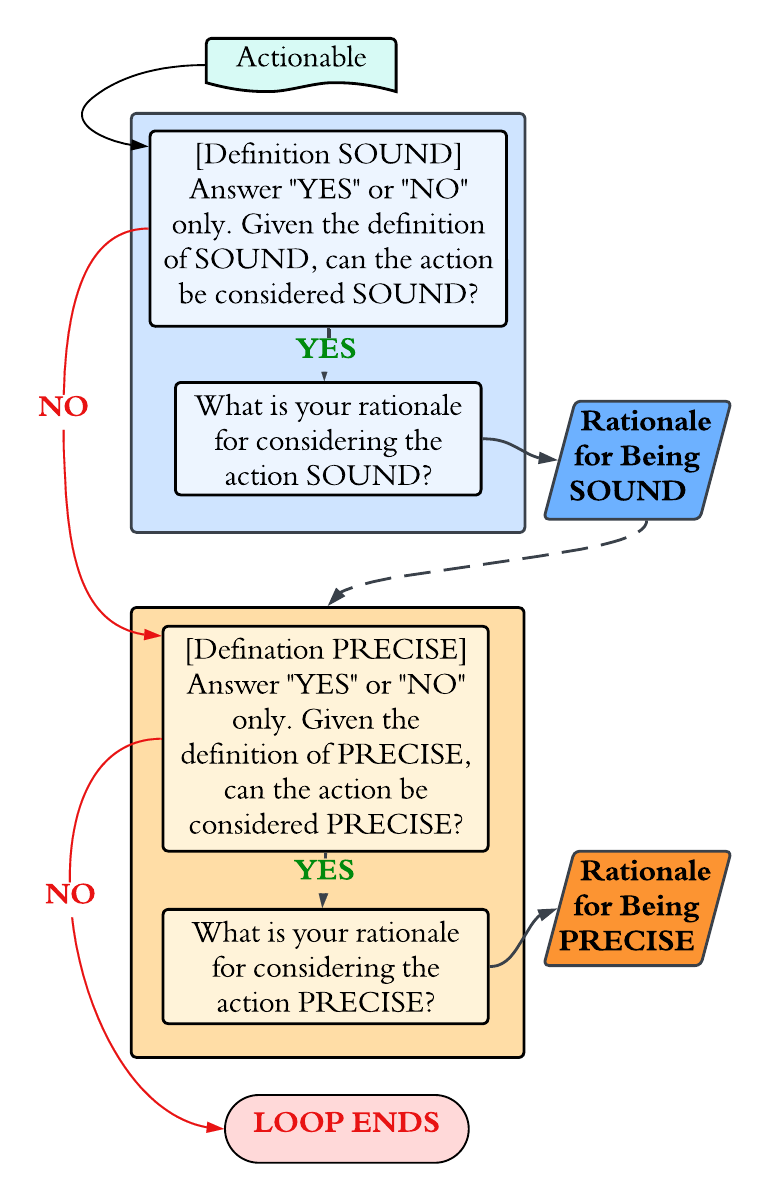}
        \caption{Flowchart of the Follow-Up Prompting Technique for ReACT Reliability Assessment}
        \label{quality}
    \end{minipage}
\end{figure*}

Based on the definition of ReACT reliability, here also, we designed a follow-up prompt to assess whether a particular ReACT is Sound and Precise. The flowchart for this reliability analysis is shown in Fig.~\ref{quality}. In this phase as well, we utilized the winner LLM from task one.

\textbf{Explicit Prompt Sequence:}
\begin{enumerate}
    \item \textbf{Soundness Assessment:} First, we provide the model with the definition of ``SOUND'' (as specified in Section~\ref{sec:llm_metrics}): ``A ReACT is sound if it makes logical sense, has no contradictions, and all parts of the recommendation work together consistently.''
    
    Then we ask: ``Answer YES or NO only. Given the definition of SOUND, can the following action be considered SOUND: [ReACT text]?''
    \begin{itemize}
        \item If YES $\rightarrow$ ``What is your rationale for considering the action SOUND?''
        \item If NO $\rightarrow$ ``What is your rationale for considering the action UNSOUND?''
    \end{itemize}
    
    \item \textbf{Preciseness Assessment:} Next, we present the definition of ``PRECISE'': ``A ReACT is precise if it is clear, easy to follow, specific, and leaves no room for confusion.''
    
    Then we ask: ``Answer YES or NO only. Given the definition of PRECISE, can the following action be considered PRECISE: [ReACT text]?''
    \begin{itemize}
        \item If YES $\rightarrow$ ``What is your rationale for considering the action PRECISE?''
        \item If NO $\rightarrow$ ``What is your rationale for considering the action IMPRECISE?''
    \end{itemize}
\end{enumerate}


This follow-up prompting process for ReACT quality assessment ensures that each ReACT meets our defined quality standards.

\begin{figure*}[]
\includegraphics[width = \linewidth, height=8.5cm]{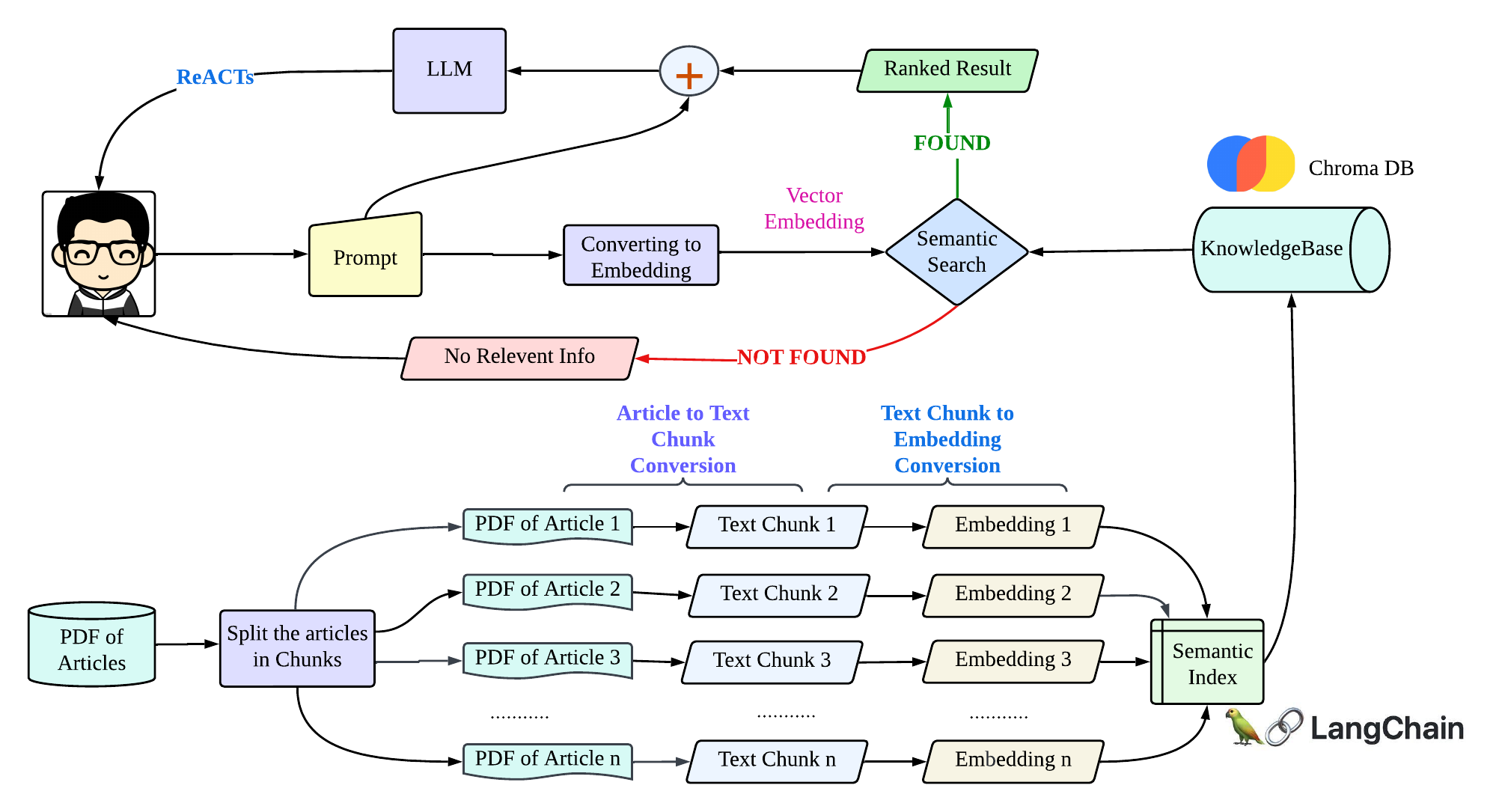}
\caption{Architecture of the proposed RAG pipeline (Adapted from EvidenceBot \cite{khan2025evidencebot})}
\label{fig:RAGPipeline}
\end{figure*}
\textbf{}

\subsection{Development of the LLM Pipeline}
\label{sec:rag_pipeline}

This section details the development process of the LLM pipeline used in our study. For designing the pipeline, we considered that LLMs can hallucinate and provide factually inaccurate information if unaware of the context of the question~\cite{22_perkovic2024hallucinations}. To prevent inaccuracies, we adopted the RAG technique, which possesses a knowledge base and can pass the required knowledge/context to the model along with the prompt~\cite{90_gao2023retrieval}. Consequently, the RAG technique helps the model understand the context of the question being asked~\cite{23_chen2024benchmarking}.

\textbf{Why RAG for Per-Article Processing?} While our pipeline processes one article at a time, RAG provides several advantages over directly feeding PDF text: (i) \emph{standardized parsing}: consistent extraction and cleaning of PDF content across all 829 articles; (ii) \emph{context injection}: seamless integration of full article context with prompts during both derivation and follow-up phases; (iii) \emph{reproducibility}: embedding-based retrieval ensures identical context formatting across experiments; and (iv) \emph{extensibility}: the same pipeline can be adapted for multi-article queries in future work. We developed a local RAG-based pipeline where we placed the articles in the knowledge base and prompted the models about the given papers.

The architecture of the proposed pipeline is illustrated in Fig.~\ref{fig:RAGPipeline}. The system accepts PDF documents, which are processed using Python's \texttt{PyPDF2} library~\cite{56_PyPDF2} to extract raw text and create text chunks. Unlike traditional RAG systems that divide documents into smaller chunks, our pipeline creates a \emph{single text chunk per article}. This approach ensures that when the retrieval process returns a text chunk, it contains the entire article rather than a subset of the article. This chunking process ensures that the model perceives the context of the entire article before deriving ReACTs.

The text chunks undergo a cleaning process where the ``ACKNOWLEDGEMENT'' and ``REFERENCE'' sections are removed as they are irrelevant for ReACT derivation. The cleaned text chunks are then converted into embeddings using a fine-tuned embedding model called \textit{instructor-large}~\cite{72_huggingface_instructor_large}. This model, proposed by Su et al.~\cite{48_embedding}, generates text embeddings tailored to various downstream tasks and achieved an average improvement of 3.4\% across 70 evaluation tasks. The embedding dimensionality is 768 (matching BERT-base architecture).

Next, the pipeline uses \texttt{LangChain}~\cite{57_LangChain} to create semantic indexes for each embedding, which enables efficient retrieval and ranking of relevant information based on meaning and context rather than simple keyword matching~\cite{50_papadimitriou1998latent}. These embeddings and their semantic indexes are stored in \texttt{ChromaDB}~\cite{73_chroma_website}, which is a high-performance vector database optimized for similarity search~\cite{49_xie2023brief}.

\textbf{Retrieval Parameters:} When a user provides a query, it is converted into an embedding using the same \textit{instructor-large} model. The pipeline performs a semantic search in the vector database with the embedded query using \emph{cosine similarity} as the distance metric. We set $k = 1$ (retrieve only the single most relevant article) since our design processes one article per query. The plain text of the retrieved article is appended to the user's query and passed to the LLM model, ensuring the model has the full context of the article. The pipeline is designed for single-article processing and does not support simultaneous multi-article queries.

\subsubsection{Independence of Pipeline Stages}
Although we use the same model across all three pipeline stages, each stage operates as an independent, stateless function call. In the extraction stage, the model receives article text and outputs candidate ReACTs. In the refinement stage, the model receives article text plus a ReACT statement (with no memory that it generated this ReACT) and validates its presence while extracting impact and evidence. In the assessment stage, the model receives only the ReACT text and quality definitions (with no context about prior stages) and classifies soundness and preciseness. 

This design prevents explicit confirmation bias—the model cannot ``defend'' its own outputs because it doesn't know they are its own. However, it does not eliminate systematic model-specific biases: if the winner model consistently interprets certain linguistic patterns as actionable or precise due to its training, these tendencies will manifest across all stages. We address this through human validation (Section~\ref{sec:sanity_check}).

\subsection{Sanity Check}
\label{sec:sanity_check}
In this step, we manually evaluated the performance of the winner LLM from task one in refining and assessing ReACTs (tasks 2 and 3). After the accomplishment of the tasks, 100 ReACTs were randomly selected from the ReACT dataset for analysis. Two independent annotators were tasked with this process. Given the full text of the articles, they were asked to evaluate each ReACT by responding in a YES/NO format for the evaluation metrics \textit{(sound, precise, impact-present, and evidence-present)}.

For the metric-\textit{sound} and \textit{precise}, agreeing with the LLM's verdict means LLM has performed the classification correctly, while disagreeing means LLM was not correct in the classification. While, for the metric \textit{impact-present} and \textit{evidence-present}, the annotators either agreed or disagreed with the LLM’s extraction. For instance, if a ReACT contained an explicit impact/evidence, agreeing indicated that the LLM had correctly identified the impact/evidence; while disagreeing suggested that the extracted impact/evidence was incorrect. Conversely, for a ReACT with no explicit impact/evidence mentioned, agreeing indicated that the LLM had correctly recognized the absence of impact, while disagreeing meant that the impact/evidence was actually present in the article, but the LLM failed to extract it.

To measure the consistency of the annotation, the Inter-Annotator Agreement (IAA) \cite{84_artstein2017inter} score was used. After the independent annotations, the reviewers discussed the annotations with each other and resolved the disagreements through discussion. After the disagreements were resolved, we got the final human labeling/gold annotation, which we compared with the LLM's actual response for performing the sanity check.

\subsection{Categorization of ReACTs}
\label{sec:categorization}

To provide OSS project-specific ReACTs to the developer community, we categorized them into eight distinct categories: a) \textit{New Contributor Onboarding and Involvement}; b) \textit{Code Standards and Maintainability}; c) \textit{Automated Testing and Quality Assurance}; d) \textit{Community Collaboration and Engagement}; e) \textit{Documentation Practices}; f) \textit{Project Management and Governance}; g) \textit{Security Best Practices and Legal Compliance}; and h) \textit{CI/CD and DevOps Automation}. These categories were established based on an in-depth discussion and careful consideration of the specific needs of OSS projects. The categories' definitions and the criteria for categorization (i.e., the conditions qualifying a ReACT for inclusion in a particular category) are provided in Section 2 of the Appendix.

Two independent researchers were tasked with assigning predefined categories to each ReACT. Although a single ReACT may be applicable to multiple categories, the annotators were instructed to assign the category most relevant to each ReACT. Like the sanity check, the quality of these annotations was assessed using the IAA score. Following the annotation process, both researchers collaboratively resolved any disagreements through discussion.

\subsection{Case Studies}

We conducted case studies with marginal projects to demonstrate how ReACTs can be effectively applied to support the development of OSS projects. The projects were selected from the Apache Software Foundation Incubator (ASFI)~\cite{75_apache_incubator}. Marginal projects refer to those that either showed initial promise but ultimately failed or those that started poorly but later succeeded.

ASFI~\cite{75_apache_incubator} serves as a gateway for OSS projects aiming to join the ASF ecosystem~\cite{76_apache}. In ASFI, projects receive mentorship from senior members on the ``Apache Way'' so they can mature and align with Apache's principles of community-driven development and open governance. The incubation process typically spans one to three years~\cite{76_apache}. Near the end of incubation, a project faces one of three possible outcomes: graduation, continuation, or retirement, decided by the project members and a Project Management Committee (PMC). The desired outcome of a project is graduation, which is only achieved if a project fulfills all ASF goals. However, if a project is not yet ready for graduation but shows promise, it may continue in the Incubator for an extended period. On the contrary, if a project fails to make sufficient progress or loses community support, it may be retired from the Incubator. 

We utilized project activity data from the APEX (Apache Project EXplorer) tool, developed by Ramchandran et al.~\cite{78_ramchandran2022exploring}. APEX is an open-source web application dashboard tool designed to monitor and explore the sustainability trajectories of 273 ASFI projects. It features an AI-based sustainability forecasting model, interactive visualizations of social and technical networks, and detailed manifestation of project activity data. In the tool, the social networks denote the communication among developers in the Apache mailing list~\cite{92_apacheMailingLists}, while technical networks represent developers' code contributions to different file types. The activity data of the tool include month-wise social and technical network, number of emails, commits, contributors, etc. The tool employs an LSTM model trained on 18 socio-technical features to generate month-by-month graduation probabilities. The graduation probability at month \textit{n} reflects the model's confidence that the project will graduate, based on data from months 0 to \textit{n-1}. If a project's graduation probability rises above 0.5, it indicates an upturn, while a drop below 0.5 signals a downturn. Using APEX, users can explore project data over time, compare different projects, and analyze sustainability turning points. The tool can be publicly accessed at \href{https://ossustain.github.io/APEX/}{this link}.

\section{RESULTS AND DISCUSSION}

\subsection{$\textbf{RQ}_{\textcolor{red}{1}}$: Can we use LLMs to effectively derive evidence-based actionable recommendations from the scientific literature?}

\textbf{Summary:} Among four LLMs and three prompting techniques evaluated on our small labeled set of 10 papers (42 annotated ReACTs), Mixtral-8x7B with Chain-of-Thought prompting achieved the highest scores across all metrics (BLEU-4: 0.66, ROUGE-L: 0.49, BERTScore: 0.75, METEOR: 0.56), substantially outperforming other combinations and establishing it as the best available model-prompt pair for ReACT extraction in our experimental setting.

We evaluated four different LLMs (Llama3-8B, Mixtral-8x7B, MistralLite-7B, and Mistral-Nemo-12B) and three distinct prompting techniques (Zero-shot, CoT, and Reason+Action) for deriving ReACTs from articles. Performance was assessed using four metrics (BLEU-4, ROUGE-L, BERTScore, and METEOR) and delineated in Table~\ref{table:LLMEvaluation}. As discussed in Section~\ref{sec:llm_metrics}, these metrics are weak proxies for true ReACT quality due to paraphrasing variability, but serve to identify the relatively best-performing model-prompt combination on our labeled set.

\textit{Mixtral-8x7B} paired with the \textit{CoT} prompting technique performed best among the evaluated configurations, substantially outperforming other combinations across all metrics. This pairing demonstrated a strong ability to balance lexical overlap and semantic depth, making it the most effective configuration among those tested for ReACT derivation. While \textit{MistralLite-7B} showed mixed results, its performance was inconsistent across prompting techniques, with \textit{Reason+Action} performing slightly better but not achieving the consistency seen with \textit{Mixtral-8x7B}.

Conversely, \textit{Llama3-8B} consistently performed poorly across all metrics and prompting techniques. Its low scores indicate that this model struggles with both lexical and semantic tasks in this context. \textit{Mistral-Nemo-12B}, despite its larger size (12B parameters), also showed disappointing results regardless of the prompt used, suggesting that parameter count alone does not guarantee better performance for specialized extraction tasks.

Given the small size and author-annotated nature of the labeled set, this comparison should be interpreted as a pragmatic model-selection heuristic, not a definitive ranking of LLM capabilities

\vspace{0.25cm}
\fbox{%
  \centering
  \parbox{0.93\linewidth}{
    \textbf{$\textbf{RQ}_{\textcolor{red}{1}}$ Findings:} LLMs can be used to derive actionable recommendations from scientific articles. Among the evaluated models and prompting techniques tested on our small labeled set (10 papers, 42 ReACTs), the combination of \textit{Mixtral-8x7B} with the \textit{Chain-of-Thought} prompting method achieved the best performance by a large margin across all metrics, establishing it as our selected configuration for full-scale extraction.
  } 
}
\vspace{0.25cm}

\begin{table*}[t]
\centering
\caption{Performance Analysis of ReACT Derivations Across Different LLMs and Prompting Techniques}
\label{table:LLMEvaluation}

\begin{tabular}{llcccc}
\hline
\textbf{Model} & 
\textbf{\begin{tabular}[c]{@{}l@{}}Prompt \\ Technique\end{tabular}} 
& \textbf{BLEU-4} & \textbf{METEOR} & \textbf{ROUGE-L} & \textbf{BERT} \\
\hline

\multirow{3}{*}{\textit{Llama3-8b}} 
& Zero Shot        & 0.04 & 0.20 & 0.09 & 0.46 \\
& Chain-of-Thought & 0.06 & 0.15 & 0.07 & 0.46 \\
& Reason+Action    & 0.08 & 0.16 & 0.09 & 0.47 \\
\hline
 
\multirow{3}{*}{\textit{Mixtral-8x7b}} 
& Zero Shot        & 0.38 & 0.38 & 0.32 & 0.62 \\
& Chain-of-Thought & \textbf{0.66} & \textbf{0.56} & \textbf{0.49} & \textbf{0.75} \\
& Reason+Action    & 0.14 & 0.17 & 0.10 & 0.46 \\
\hline
 
\multirow{3}{*}{\textit{MistralLite-7b}} 
& Zero Shot        & 0.44 & 0.22 & 0.23 & 0.52 \\
& Chain-of-Thought & 0.31 & 0.11 & 0.12 & 0.46 \\
& Reason+Action    & 0.49 & 0.30 & 0.27 & 0.57 \\
\hline

\multirow{3}{*}{\textit{Mistral Nemo-12B}} 
& Zero Shot        & 0.10 & 0.14 & 0.08 & 0.45 \\
& Chain-of-Thought & 0.12 & 0.14 & 0.08 & 0.43 \\
& Reason+Action    & 0.09 & 0.14 & 0.07 & 0.43 \\
\hline

\end{tabular}
\end{table*}

\subsection{$\textbf{RQ}_{\textcolor{red}{2}}$: How many evidence-based actionable recommendations can LLMs derive from scientific literature published in top SE venues (ICSE and FSE) in the domain of OSS, and what is the reliability of those derived actionables?}

\textbf{Summary:} From 829 ICSE/FSE papers, we extracted 2,023 candidate ReACTs. After filtering 101 hallucinations through follow-up prompting, 1,922 validated ReACTs remained. Of these, 1,673 (87\%) include explicit impact statements, 1,458 (76\%) include empirical evidence, 1,901 (99\%) are sound, and 1,697 (88\%) are precise. Overall, 1,312 ReACTs (68\%) meet all quality criteria (sound, precise, with both impact and evidence), constituting our complete ReACT catalog. Human validation on a stratified sample of 100 ReACTs showed 87–100\% agreement with LLM classifications across metrics.

Leveraging the findings from ${RQ}_{\textcolor{black}{1}}$, we used the combination \textit{Mixtral-8x7B} with \textit{CoT} prompting to extract actionables from the entire dataset of 829 articles. We successfully derived ReACTs from 474 articles (57\%), while the remaining 355 articles (43\%) did not produce any ReACTs—either because they did not focus on actionable practices or because their findings were too abstract for concrete extraction. The model's self-confidence score for ReACT derivation averaged 0.8, with a standard deviation of 0.1436. In total, this initial layer of analysis resulted in the extraction of 2,023 ReACTs, or approximately 4.3 per article on average (calculated over the 474 articles that yielded ReACTs).

Following the initial extraction, we implemented our adapted two-phase second-layer prompting. The first phase involved a series of targeted questions to the model, designed to ascertain three key aspects: (1) whether the ReACT was explicitly mentioned in the article, (2) the impact of the actionable item mentioned in the article, and (3) the stated evidence supporting the ReACT in the article. During this phase, \textit{Mixtral-8x7B} was utilized as the LLM model, as it emerged as the top-performing LLM in our study based on the findings from ${RQ}_{\textcolor{black}{1}}$. This step identified 101 ReACTs (5\% of initial extractions) that were not actually mentioned in the articles. These were classified as hallucinated ReACTs and were consequently removed from the dataset prior to further analysis. This filtering process resulted in a refined set of 1,922 ReACTs for subsequent examination.

The next phase in the second layer of prompting was to supplement each of the remaining ReACTs with impact and evidence information. The model successfully identified explicitly mentioned impacts for 1,673 ReACTs (87\%), while no explicit impact was found for 249 ReACTs (13\%). It is worth noting that the impact of some actionable items is implicit in the action itself (e.g., ``Implement automated testing'' implicitly improves reliability), which explains why the authors of the scientific articles do not always explicitly state the impact of the actionable in their articles. In terms of evidence, the model identified explicit empirical evidence for 1,458 ReACTs (76\%) out of the 1,922 analyzed. The lower evidence extraction rate compared to impact suggests that evidence statements are more challenging for the LLM to identify—possibly because evidence often appears in scattered form across Results and Discussion sections rather than being localized near the actionable statement. Notably, the model found 237 actionable items (12\%) for which there was no explicit mention of either impact or evidence in the source articles.

In the final phase of our analysis, we employed the top-performing LLM from our study, \textit{Mixtral-8x7B}, to evaluate the ReACTs based on their soundness and preciseness characteristics. This process involved passing each actionable through the model and posing a series of questions to determine whether each action item met the Sound and Precise criteria. The results of this analysis revealed that out of the 1,922 ReACTs, only 21 (1.1\%) were classified as Unsound, while 1,901 (98.9\%) were classified as Sound. Additionally, 226 ReACTs (11.8\%) were categorized as Imprecise, with 1,697 (88.2\%) classified as Precise. Interestingly, 8 ReACTs were found to be both Unsound and Imprecise, representing cases where logical inconsistencies coincided with vague or unclear phrasing. Overall, we identified 1,312 ReACTs (68.3\% of validated ReACTs) as \textbf{complete}, meeting all criteria: SOUND, PRECISE, and having explicit mentions of both impact and evidence.

\subsubsection{Human Validation and Sanity Check}

To validate the efficacy of our study's methodology, we conducted a manual evaluation of the ReACTs as described in Section~\ref{sec:sanity_check}. This process involved selecting 100 ReACTs via stratified random sampling across the entire set of ReACTs. Two independent evaluators participated in this validation process, and their consistency was measured by IAA. The IAA between the annotators were: 95 for impact, 94 for evidence, 98 for soundness, and 96 for preciseness, indicating strong consistency in human judgment.

Next, the annotators resolved their disagreements through discussion, and we obtained the final human-labeled gold annotation. We then compared the gold annotation with the LLM's original response. We observed a match rate of 98\% for \textit{impact-present} and 87\% for \textit{evidence-present}. The lower match rate for evidence (87\% vs. 98\% for impact) confirms that accurately extracting ReACT evidence is more challenging for the LLM than extracting impact—likely because evidence statements are more scattered and varied in form across papers. Soundness achieved a 96\% match, while preciseness achieved a 98\% match. The near-perfect agreement on impact and preciseness—underscores the robustness and efficacy of our two-layer refinement and reliability assessment approach in extracting and analyzing ReACTs from academic literature.

It is important to clarify what these high soundness and preciseness rates do and do not indicate. Soundness, as defined in this study, is a \textit{logical coherence check}: a ReACT is considered sound if its parts are internally consistent and free of explicit contradictions, not if the recommended action is guaranteed to be effective in practice. Likewise, a high preciseness score indicates that the recommendation is clearly articulated, specific, and easy to follow, not that it is universally applicable or sufficient on its own to ensure project sustainability. Preciseness reflects linguistic clarity and operational specificity, rather than contextual fit or empirical strength beyond what is reported in the source study.

\vspace{0.4cm}
\fbox{%
  \centering
  \parbox{0.93\linewidth}{
    \textbf{$\textbf{RQ}_{\textcolor{red}{2}}$ Findings:} We identified 1,922 validated ReACTs from 829 ICSE/FSE papers. Of these, 1,673 (87\%) were supported by explicit impact, 1,458 (76\%) by explicit evidence, 1,901 (99\%) classified as SOUND, and 1,697 (88\%) as PRECISE. Consequently, 1,312 ReACTs (68\%) met all criteria (SOUND, PRECISE, with both impact and evidence), constituting our complete ReACT catalog. Human validation on 100 stratified-sampled ReACTs showed 87–100\% agreement with LLM classifications.
  }%
}

\subsection{$\textbf{RQ}_{\textcolor{red}{3}}$: How can the actionables be presented meaningfully in practice?}

\textbf{Summary:} We organized 1,922 ReACTs into eight practice-oriented categories and demonstrated their practical application through case studies of two ASFI projects. The case studies show how maintainers can use APEX to identify sustainability turning points, select relevant ReACTs from appropriate categories, and implement evidence-based interventions during critical project phases.

To answer this RQ, we first categorized the ReACTs into practice-oriented categories, then performed case studies on projects from the ASFI using the APEX tool to demonstrate how ReACTs can guide real-world OSS projects toward sustainability.

\subsubsection{Categorization of ReACTs}

We manually categorized the ReACTs into eight predefined categories as described in Section~\ref{sec:categorization}. The consistency of the annotation was evaluated with IAA, which yielded a score of 0.84, indicating strong agreement. Next, for each category, we gathered statistical data to assess both the support for the ReACTs and the quality of that support. A summary of these statistics is provided in Table~\ref{table:ReACTCategories}. 

Several notable patterns emerge from this categorization:

\begin{itemize}
    \item \textbf{Distribution:} The majority of ReACTs were assigned to the category \textit{Automated Testing and Quality Assurance} (n=607, 31.6\%), while the fewest were placed in the \textit{New Contributor Onboarding and Involvement} category (n=60, 3.1\%). This distribution reflects the historical focus of OSS research on technical quality practices over community-building practices.
    
    \item \textbf{Completeness Variability:} The percentage of Complete ReACTs (those meeting all quality criteria) ranged from 54\% (\textit{Community Collaboration and Engagement} and \textit{Documentation Practices}) to 85\% (\textit{CI/CD and DevOps Automation}). This substantial variation suggests that research in DevOps and automation tends to provide more explicit, well-evidenced recommendations, while community-oriented research often presents findings that are less directly actionable or backed by less concrete empirical evidence. This may reflect the relative maturity of automated practices research versus socio-technical community research, or it may indicate that technical practices are easier to operationalize into concrete actions.
    
    \item \textbf{Evidence Backing:} ReACTs backed by evidence ranged from 60\% (\textit{Community Collaboration and Engagement}) to 95\% (\textit{CI/CD and DevOps Automation}). The lower evidence rates in community-oriented categories may reflect the methodological challenges of empirically validating social interventions in volunteer OSS contexts, whereas technical practices (testing, CI/CD) can be more readily evaluated through quantitative metrics (build times, defect rates, etc.).
\end{itemize}

Across all categories, the percentage of Sound ReACTs was consistently high (98–100\%), indicating that nearly all extracted recommendations are logically coherent. PRECISE ReACTs ranged from 83\% to 91\%, suggesting that while most ReACTs provide clear guidance, approximately 10–15\% remain somewhat vague or require additional context for implementation. The variability in LLM confidence scores was minimal, consistently ranging between 77\% and 81\%, indicating stable model confidence across different practice areas.

\begin{table*}[]
\centering
\caption{Summary of ReACT Derivations Across Various ReACT categories; The table presents a breakdown of the total number of articles contributed to a category, ReACTs for each category, along with the percentage of those which are SOUND, PRECISE, contain empirical evidence, and have an impact. The 'Complete ReACTs' column represents ReACTs that meet all these criteria. The notation (\# (\%)) indicates both the actual count of ReACTs and the percentage relative to the total number of ReACTs in that category. The final column shows the average LLM confidence score for each category.}
\label{table:ReACTCategories}
\def\arraystretch{1.2}
\resizebox{\textwidth}{!}
{
\begin{tabular}{l|ccccccc|c}
\hline
\multicolumn{1}{c|}{\textbf{ReACT Category}}  & \textbf{\begin{tabular}[c]{@{}l@{}}\# of \\ Articles\end{tabular}} & \textbf{\begin{tabular}[c]{@{}l@{}}\# of \\ ReACTs\end{tabular}} & \textbf{ \begin{tabular}[c]{@{}l@{}}SOUND\\  ReACTs\\ (\# (\%))\end{tabular}} & \textbf{\begin{tabular}[c]{@{}l@{}}PRECISE \\ ReACTs\\ (\# (\%)) \end{tabular}} & \textbf{\begin{tabular}[c]{@{}l@{}} ReACTs\\ with \\ Impact\\ (\# (\%)) \end{tabular}} & \textbf{\begin{tabular}[c]{@{}l@{}} ReACTs\\ with \\ Evidence\\ (\# (\%))\end{tabular}} & \textbf{\begin{tabular}[c]{@{}l@{}}Complete\\ ReACTs \\(\# (\%))\end{tabular}} & \textbf{\begin{tabular}[c]{@{}l@{}}LLM\\ Confidence\\ Score (\%)\end{tabular}} \\ \hline

\textit{\begin{tabular}[c]{@{}l@{}} New Contributor Onboarding \\and Involvement \end{tabular}}  & 43 & 60 & 59 (98.33\%) & 54 (90.00\%) & 50 (83.33\%) & 46 (76.67\%) & 42 (70.00\%) & 79.92 \\ \hdashline

\textit{\begin{tabular}[c]{@{}l@{}} Code Standards \\and Maintainability\end{tabular}}  & 228 & 394 & 388 (98.48\%) & 358 (90.86\%) & 354 (89.85\%) & 310 (78.68\%) & 288 (73.10\%) & 81.11 \\ \hdashline

\textit{\begin{tabular}[c]{@{}l@{}} Automated Testing \\and Quality Assurance \end{tabular}}  & 309 & 607 & 602 (99.18\%) & 532 (87.64\%) & 531 (87.48\%) & 481 (79.24\%) & 431 (71.00\%) & 80.83 \\ \hdashline

\textit{\begin{tabular}[c]{@{}l@{}} Community Collaboration \\and Engagement \end{tabular} } & 139 & 242 & 239 (98.76\%) & 212 (87.60\%) & 197 (81.40\%) & 146 (60.33\%) & 131 (54.13\%) & 78.47 \\ \hdashline

\textit{\begin{tabular}[c]{@{}l@{}} Documentation \\ Practices \end{tabular}}  & 133 & 173 & 169 (97.69\%) & 144 (83.24\%) & 146 (84.39\%) & 109 (63.01\%) & 96 (55.49\%) & 78.27 \\ \hdashline

\textit{\begin{tabular}[c]{@{}l@{}} Project Management \\and Governance\end{tabular}}  & 94 & 154 & 153 (99.35\%) & 132 (85.71\%) & 126 (81.82\%) & 105 (68.18\%) & 92 (59.74\%) & 77.32 \\ \hdashline

\textit{\begin{tabular}[c]{@{}l@{}} Security Best Practices \\and Legal Compliance\end{tabular}}  & 81 & 155 & 155 (100.00\%) & 138 (89.03\%) & 140 (90.32\%) & 131 (84.52\%) & 115 (74.19\%) & 80.35 \\ \hdashline

\textit{\begin{tabular}[c]{@{}l@{}} DevOps Automation \\and CI/CD\end{tabular} } &  87 & 137 & 136 (99.27\%) & 126 (91.97\%) & 129 (94.16\%) & 130 (94.89\%) & 117 (85.40\%) & 80.69 \\
\hline
\end{tabular}
}
\end{table*}

\subsubsection{Case Studies: Operationalizing ReACTs with APEX}

\begin{figure*}[]
\begin{minipage}{.90\textwidth}
  \centering
  \includegraphics[width=\linewidth,height=7.5cm]{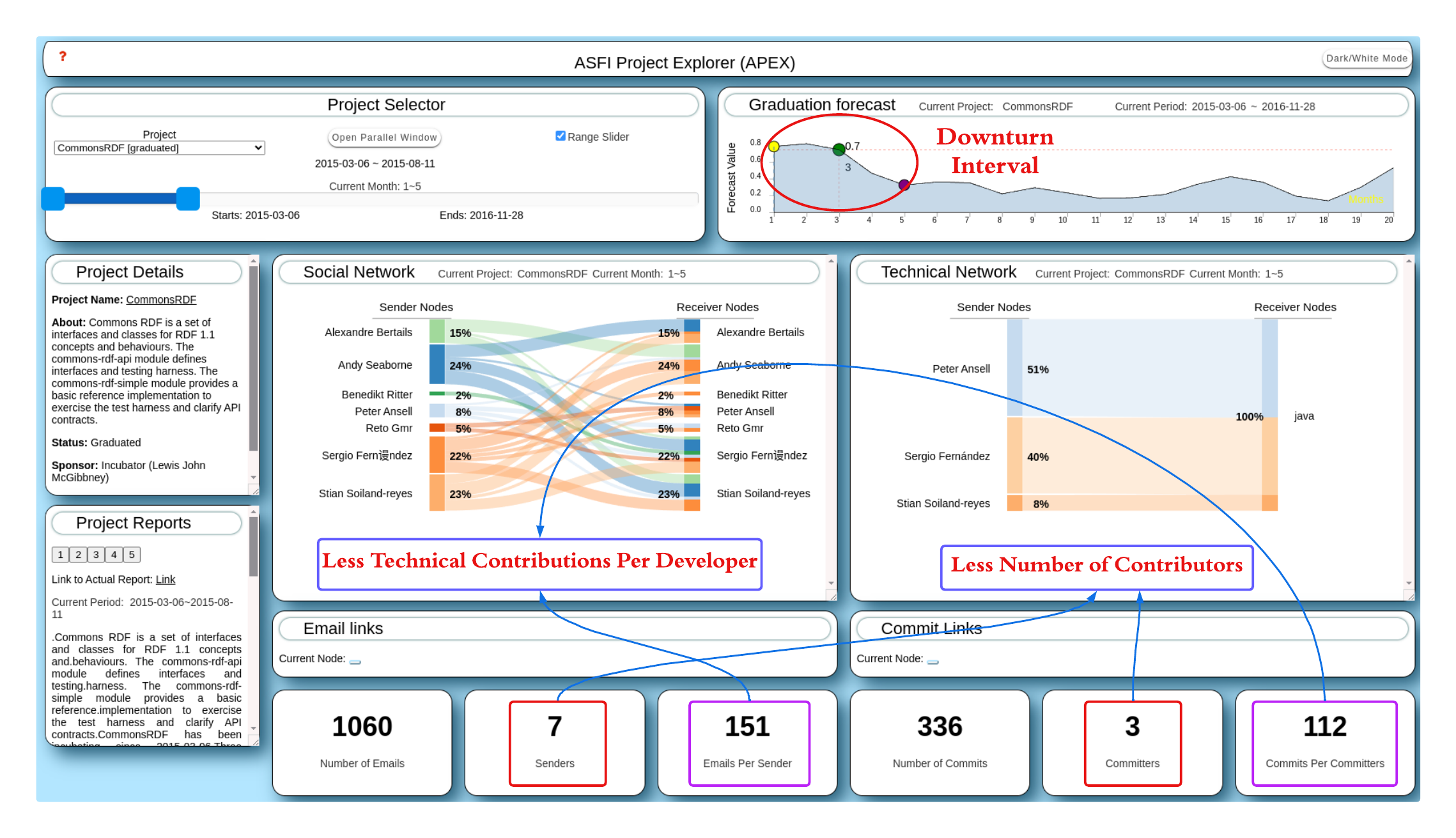}
  \subcaption{}
\end{minipage}%

\begin{minipage}{.90\textwidth}
  \centering
  \includegraphics[width=\linewidth,height=7.5cm]{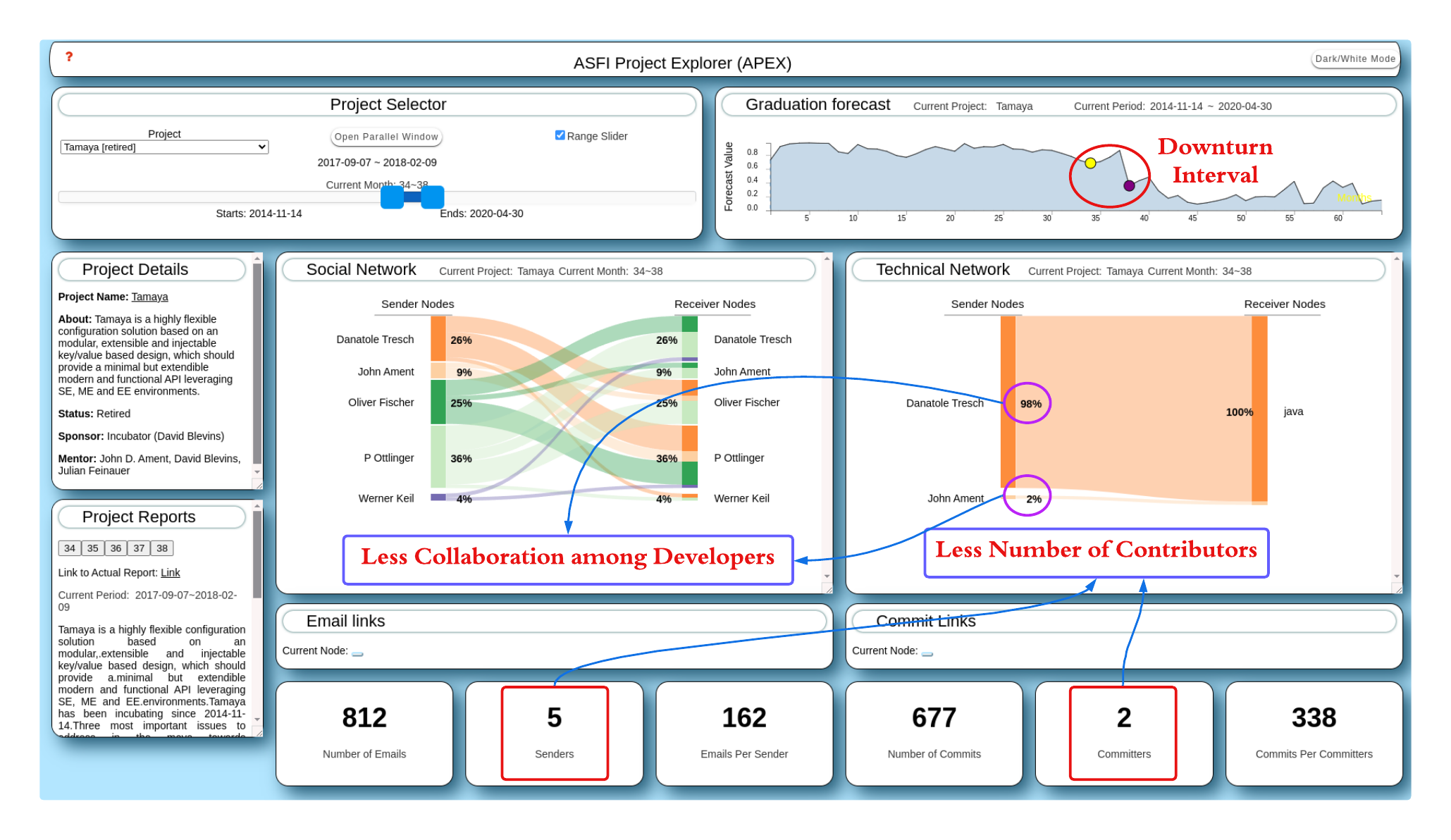}
  \subcaption{}
\end{minipage}%

\caption{Metrics for two marginal projects from ASFI: (a) graduated project \textit{CommonsRDF} which almost failed, while (b) retired project \textit{Tamaya} which almost succeeded. The screenshots are adopted from the APEX tool \cite{52_APEX}. The figures include annotations (in red) that analyze the factors contributing to the downturn.}
\label{fig:Case_Studies}
\end{figure*}

We selected two projects for case studies to \textit{illustrate how ReACTs can be operationalized in practice}. These case studies are intended as exploratory demonstrations of how practitioners could use ReACTs to interpret project signals and identify relevant evidence-based actions. The selected projects include one that faced an initial downturn in its lifecycle but later graduated (\textit{CommonsRDF}) and another that showed good promise in the beginning but later retired (\textit{Tamaya}). Project activity data and graduation forecasts of these projects are shown in Fig.~\ref{fig:Case_Studies} (obtained from the APEX tool). Both projects faced significant downturns in their lifecycles (see Graduation Forecast panels). For \textit{CommonsRDF}, a significant downturn can be observed in month 5, while \textit{Tamaya} experienced a downturn in month 39.

\textbf{Maintainer Workflow for Using ReACTs:} We describe a workflow that project maintainers can follow to select and implement ReACTs based on sustainability signals:

\paragraph{Step 1: Identify Turning Points.} Using APEX's graduation forecast visualization, maintainers identify months where the sustainability probability drops below 0.5 (downturn) or exhibits sustained decline. We selected downturn intervals for both projects to analyze the causes of decline. The downturn interval includes two months before and two months after the turning point. Thus, the downturn interval for \textit{CommonsRDF} spans months 1–5, while for \textit{Tamaya} it spans months 37–41.

\paragraph{Step 2: Analyze Socio-Technical Signals.} The APEX tool has a range slider feature that allows users to select an interval. Selecting an interval enables the tool to show aggregated project activity data during that interval—including contributor counts, commit frequencies, communication patterns (social network), and code contribution patterns (technical network). Consequently, we selected the stated downturn intervals for the projects and examined their socio-technical indicators.

\paragraph{Step 3: Map Signals to ReACT Categories.} Based on the observed deficits, maintainers can select relevant ReACT categories. For example, low contributor counts suggest \textit{New Contributor Onboarding and Involvement} ReACTs; high code complexity or low contributions per developer suggest \textit{Code Standards and Maintainability} ReACTs; minimal collaboration suggests \textit{Community Collaboration and Engagement} ReACTs.

\paragraph{Step 4: Select Evidence-Based ReACTs.} From the selected categories, maintainers review ReACTs and choose those most applicable to their project context, prioritizing Complete ReACTs (those with both impact and evidence) to maximize confidence in the intervention.

\paragraph{Step 5: Implement and Monitor.} After implementing selected ReACTs, maintainers can monitor subsequent metrics (months $n+1$, $n+2$, etc.) of the project to assess whether socio-technical indicators improve and whether the graduation forecast trajectory recovers.

\textbf{Case Study 1: CommonsRDF}

For the project \textit{CommonsRDF}, APEX data during the downturn period (months 1–5) revealed that only seven individuals contributed to the social network, while three contributed to the technical network. This suggests the project is in need of recruiting new developers to mitigate its struggling condition. Additionally, there were fewer technical contributions per developer during this time (averaging 8 commits per developer compared to 15 in healthy projects), highlighting the need for improved coding standards to reduce the challenges developers face when contributing to the project's codebase.

While the necessary interventions (recruit contributors, improve code quality) are clear from APEX signals, implementing them is not straightforward. Recruiting new contributors in an OSS project is inherently challenging due to the voluntary nature of OSS participation. Likewise, enhancing code quality is a complex and gradual process. This is where ReACTs provide actionable leverage. Actionables from the categories \textit{New Contributor Onboarding and Involvement} and \textit{Code Standards and Maintainability} offer evidence-based guidelines that developers can follow to ensure long-term accomplishment of the required tasks.

Some of the ReACTs from our compiled catalog that could help the \textit{CommonsRDF} project include:

\begin{enumerate}
    \item \textbf{Encourage mentors to provide feedback on pull requests (PRs) by acknowledging their efforts and contributions} (\textit{New Contributor Onboarding and Involvement})
    \begin{itemize}
        \item \textit{Impact:} Increases newcomer retention and accelerates their path to becoming regular contributors
        \item \textit{Evidence:} Mixed-methods study of 12 Apache projects found that acknowledged newcomers were 2.3× more likely to make subsequent contributions
        \item \textit{Source:} Feng et al.~\cite{81_feng2022case}
    \end{itemize}
    
    \item \textbf{Consider synchronizing branches before applying refactorings that are likely to be incompatible with changes made in parallel} (\textit{Code Standards and Maintainability})
    \begin{itemize}
        \item \textit{Impact:} Reduces merge conflicts and integration overhead, making it easier for developers to contribute
        \item \textit{Evidence:} Analysis of 10,000+ merge scenarios in 30 projects showed that pre-synchronization reduced merge conflicts by 41\%
        \item \textit{Source:} Oliveira et al.~\cite{82_oliveira2023code}
    \end{itemize}
    
\end{enumerate}

\textbf{Expected Outcome:} By implementing these ReACTs, \textit{CommonsRDF} maintainers would have evidence-based strategies to (a) increase newcomer retention through acknowledged mentorship, and (b) reduce technical friction via better branching practices. Subsequent monitoring of the project (months 6–10) would reveal whether contributor counts increased and whether the graduation forecast trajectory improved.

\textbf{Case Study 2: Tamaya}

Similar challenges are evident for the project \textit{Tamaya}, with a limited number of active contributors during its downturn phase (months 37–41): five in the social network and two in the technical network. Notably, there is minimal collaboration among developers on the technical side. Of the two technically active contributors, one individual is responsible for 98\% of the commits, while the other contributes only 2\%. This extreme concentration indicates a lack of distributed ownership and knowledge sharing—a critical vulnerability for project sustainability.

Therefore, during \textit{Tamaya}'s downturn phase, implementing ReACTs from the categories \textit{Community Collaboration and Engagement} and \textit{New Contributor Onboarding and Involvement} could help redirect its trajectory. These evidence-based recommendations can support the project in overcoming its current challenges. Specific ReACTs that \textit{Tamaya} could have adopted during this phase include:

\begin{enumerate}
    \item \textbf{Develop a clear code of conduct that outlines expectations for behavior within the community} (\textit{New Contributor Onboarding and Involvement})
    \begin{itemize}
        \item \textit{Impact:} Establishes safe, inclusive environment that attracts diverse contributors
        \item \textit{Evidence:} Analysis of GitHub projects showed that adoption of codes of conduct correlated with 15\% increase in contributor diversity and 22\% reduction in negative interactions
        \item \textit{Source:} Zhao et al.~\cite{79_zhao2023distribution}
    \end{itemize}
    
    \item \textbf{Incorporate gamification approaches, such as awarding badges or points, to incentivize community-oriented initiatives} (\textit{Community Collaboration and Engagement})
    \begin{itemize}
        \item \textit{Impact:} Increases engagement and motivates contributions through recognition mechanisms
        \item \textit{Evidence:} Controlled study with 50 OSS projects found that gamification increased participation by 28\% and sustained engagement over 6+ months
        \item \textit{Source:} Miller et al.~\cite{80_miller2023we}
    \end{itemize}
    
\end{enumerate}

\textbf{Counterfactual Reflection:} Historically, \textit{Tamaya} did not implement such interventions and was eventually retired. Had maintainers used APEX to detect the downturn at month 39 and applied these ReACTs, they could have potentially addressed the knowledge concentration (98\% commits by one person) and low collaboration. Subsequent APEX monitoring would have shown whether new contributors joined, whether commit distribution improved, and whether the graduation forecast stabilized. While we cannot definitively prove these ReACTs would have saved \textit{Tamaya}, the case study illustrates how the ReACT catalog provides maintainers with concrete, evidence-based options at critical junctures—options that did not exist in a structured, accessible form before this work.

It should be noted that these case studies serve as proof-of-concept demonstrations of how ReACTs can be mapped to observable project states, rather than as evidence that the recommended actions were adopted or that they produced the described outcomes. The goal is to show that ReACTs can be meaningfully connected to real project conditions, the question of whether their implementation leads to sustained improvements remains an open and important direction for future empirical work

\vspace{0.25cm}
\fbox{%
  \centering
  \parbox{0.93\linewidth}{
    \textbf{$\textbf{RQ}_{\textcolor{red}{3}}$ Findings:} We present a systematic approach detailing when and how actionable recommendations can be implemented to steer the future trajectories of OSS projects toward success. ReACTs are organized into eight practice-oriented categories with 54–85\% completeness rates across categories. Case studies demonstrate a concrete five-step maintainer workflow: (1) identify turning points via APEX, (2) analyze socio-technical signals, (3) map signals to ReACT categories, (4) select evidence-based ReACTs, and (5) implement and monitor outcomes.
  } 
}
\vspace{0.25cm}

\section{Threats to Validity}

This study has the following limitations, each tied to specific methodological choices:

\textbf{Venue Selection and Generalizability.} We considered only articles published in top-tier SE conference venues (ICSE and FSE). This choice was motivated by their high citation impact and quality control, but it introduces potential bias: the ReACT catalog may underrepresent findings from practitioner-oriented venues (e.g., ICSE-SEIP, MSR), journal publications (e.g., IEEE Software, TOSEM, EMSE, JSS), or industry reports. Consequently, our ReACTs may emphasize academic research priorities (e.g., technical quality, testing) over practitioner concerns (e.g., governance, legal compliance, business models). Future studies should expand the corpus to include journal articles and practitioner venues to broaden coverage of non-code aspects such as governance, licensing, and community management.

\textbf{In-Context Learning vs. Fine-Tuning.} We considered only in-context learning, where LLMs perform tasks based on examples or instructions provided within the input context~\cite{91_dong2022survey}. In this approach, no fine-tuning is performed and the model tackles new tasks by leveraging patterns and information presented in the given prompt. However, LLM models can struggle with in-context learning by putting more focus on the later part of the input sequence~\cite{31_inContexProblem}, potentially biasing extraction toward conclusions and recommendations that typically appear near the end of papers. Future studies may focus on using fine-tuned models for ReACT derivation by using a labeled dataset containing the mapping between articles and ReACTs. Fine-tuning allows the model to specialize in targeted extraction tasks, improving performance on domain-specific applications while maintaining general knowledge.

\textbf{Model and Hardware Constraints.} Due to our hardware constraints (12 GB VRAM), we could not run very large open-source models, such as \textit{Grok-1} (314 billion parameters)~\cite{68_GrokOS}, \textit{Mixtral-8x22B} (141 billion parameters)~\cite{67_Mixtral8x22B}, or \textit{DBRX} (132B parameters)~\cite{66_DBRXBase}. These models have the potential to provide better results as they are trained on vast amounts of data and possess more complex architectures. Likewise, we could not consider some advanced prompting techniques, such as \textit{Tree-of-Thought (ToT)}, \textit{Few-Shot Prompting}, and \textit{CoT with Self-Consistency}, as these techniques require large-parameter models with extended context windows to generate reliable responses~\cite{40_tot1, 40_tot2, 43_fewshot}. Therefore, the potential for future research lies in exploring how large-scale parameter models with extended context windows, combined with advanced prompting techniques, perform in deriving actionable recommendations. Such models may extract more nuanced ReACTs or better disambiguate vague statements.

\textbf{Evaluation Metrics and Paraphrasing.} The evaluation metrics used for model selection, \textit{BERTScore}, \textit{METEOR}, \textit{BLEU-4}, and \textit{ROUGE-L}, focus mainly on quantitative performance computed against a single gold phrasing and may not fully capture qualitative aspects such as interpretability, robustness, and sensitivity to specific errors~\cite{45_ProblemWithMetrics}. Because valid ReACTs can be expressed in many paraphrased forms, these metrics provide only weak proxies for ReACT quality. We mitigated this limitation by using these metrics solely for relative model selection on a small labeled set, then relying on human sanity checks and inter-annotator agreement as the primary evidence of quality. Future studies could concentrate on developing qualitative metrics or human-in-the-loop evaluation frameworks that address these gaps, aiming to provide a more comprehensive evaluation of model performance that captures aspects like interpretability, robustness, and sensitivity to errors not fully covered by existing quantitative metrics.

\section{Ethics and Data Use}
All full-text PDFs were processed locally on our own server infrastructure, with no data transmitted to external cloud services or APIs. This approach ensures that proprietary or embargoed research content was not inadvertently leaked or shared. Our outputs—the ReACT catalog—consist of high-level, abstracted actionable recommendations rather than verbatim reproduction of full article text, mitigating copyright concerns. Each ReACT is explicitly linked to its source paper via citation, enabling readers to consult the original work for full context while respecting intellectual property rights. No human subjects data or personally identifiable information was involved in this study. The APEX tool used for case studies relies on publicly available Apache Software Foundation project data (mailing lists, commit logs), and no private or sensitive project information was accessed.

\section{Conclusion} This study presents a novel approach to enhancing OSS project sustainability through the application of LLMs for extracting actionable recommendations from the scientific literature. Our research demonstrates the efficacy of the \textit{Mixtral-8x7B} model with \textit{Chain-of-Thought} prompting in deriving evidence-based actionable recommendations (ReACTs). From a corpus of 829 articles, we extracted 1,922 unique ReACTs, with 1,312 meeting rigorous criteria for soundness, preciseness, and empirical support. The introduction of a systematic approach for categorizing these ReACTs into thematic groups and applying them practically, as illustrated through case studies of ASFI projects, offers a scalable and replicable method for addressing specific challenges in OSS project lifecycles. Our methodology, incorporating a two-layer prompting technique and manual validation, ensures the quality and relevance of the extracted ReACTs. While acknowledging the limitations, this study represents a significant advancement in leveraging AI technologies to address OSS sustainability. By bridging the gap between academic research and practical application, we provide a valuable resource for the OSS community and open new avenues for future research, including the exploration of long-term impacts of ReACT implementation and the development of dynamic recommendation systems based on real-time project metrics.

\section{Data Availability Statement}
The complete datasets and code for replication are available at the following link: <\href{https://zenodo.org/records/13744866}{\textit{https://zenodo.org/records/13744866}}>.
For convenient access, the final set of actionable recommendations can also be accessed via the following web link:<\href{https://docs.google.com/spreadsheets/d/e/2PACX-1vSZ_1JGbF3HS27VhBSkJjcRhxtr1FhniDxDwt-QhjlTMlwOuQ1bz29-1eGGvZgYI-vF8tRjMUMJm-5r/pubhtml}{\textit{Actionable-Set}}>

\section{ACKNOWLEDGMENTS}
The research received funding from the National Science Foundation under Grant No. 2020751 

\bibliographystyle{unsrt}
\bibliography{references}

\clearpage
\onecolumn
\appendix

\section{LLM Prompts}
\label{appendix:prompts}

This section outlines the various prompting techniques employed in the study. Here, \textit{TitleOfTheArticle} is a string variable updated based on the actual title of the article. The highlighted portion (italic) in the prompt indicates the emotional stimuli, which was added at the bottom of each prompt.

\subsection{Zero Shot}
\begin{mdframed}[]
You have been entrusted with the complete article titled \textit{TitleOfTheArticle}. Your objective is to meticulously extract actionable recommendations from this article. These recommendations should be aimed at enhancing the sustainability of open-source software projects.

\textit{Your attention to detail in this task is crucial, as its successful completion holds significant importance for my career advancement.}
\end{mdframed}

\subsection{Chain-of-Thought}
\begin{mdframed}[]
You have been given the full text of the article titled \textit{TitleOfTheArticle}. Your task is to extract actionable recommendations from the article. An actionable recommendation is a practical, evidence-based suggestion that provides specific, clear steps or instructions which, when implemented, are expected to produce tangible and positive results. These recommendations should be practical, making them easy to implement in real-world scenarios, and evidence-based, supported by data or findings from the article. They must be specific, providing clearly defined steps or instructions, and result-oriented, aimed at producing tangible and positive outcomes. These recommendations should offer concrete guidance that can be directly put into practice to achieve desired results. Adopting these actionable recommendations can help make open-source software projects more sustainable.

Let's break down the problem into the following steps:

\textbf{Step 1:} Carefully read each sentence of the article and identify recommendations or suggestions. Look for imperative sentences or phrases that give commands, make requests, or offer instructions to direct or persuade someone to perform a specific action.

\textbf{Step 2:} Extract the practical suggestions from the identified sentences and present them as concise, unambiguous statements in a list format.

\textbf{Step 3:} Review the list of recommendations and remove any duplicate items to ensure clarity and avoid redundancy.

\textbf{Step 4:} For each recommendation in the final list, assign a confidence level on a scale of 0 to 1, indicating how confident you are in the effectiveness of the recommendation for making open-source projects sustainable. Provide a brief explanation for your confidence level.

\textit{Your attention to detail in this task is crucial, as its successful completion holds significant importance for my career advancement.}
\end{mdframed}

\subsection{Reason + Action}
\begin{mdframed}[]
\textbf{Thought 1:} I have been given the article titled \textit{TitleOfTheArticle}. My task is to extract actionable recommendations from the article. An actionable recommendation is a practical, evidence-based suggestion that provides specific, clear steps or instructions which, when implemented, are expected to produce tangible and positive results. These recommendations should be practical, making them easy to implement in real-world scenarios, and evidence-based, supported by data or findings from the article. They must be specific, providing clearly defined steps or instructions, and result-oriented, aimed at producing tangible and positive outcomes. These recommendations should offer concrete guidance that can be directly put into practice to achieve desired results. Adopting these actionable recommendations can help make open-source software projects more sustainable.

\textbf{Action 1:} Carefully read each sentence of the article, focusing on identifying imperative sentences or phrases that give commands, make requests, or offer instructions to direct or persuade someone to perform a specific action.

\textbf{Observation 1:} A list of potential actionable recommendations has been identified from the article.

\textbf{Thought 2:} I need to extract the practical suggestions from the identified sentences and present them as concise, unambiguous statements.

\textbf{Action 2:} Convert the identified sentences into clear, actionable statements and compile them into a list.

\textbf{Observation 2:} A list of actionable recommendations has been compiled from the article.

\textbf{Thought 3:} To ensure clarity and avoid redundancy, I should review the list and remove any duplicate items.

\textbf{Action 3:} Review the list of recommendations and remove any duplicate or highly similar items.

\textbf{Observation 3:} The list of recommendations has been refined, removing duplicates and ensuring clarity.

\textbf{Thought 4:} I need to assess the confidence level for each recommendation to indicate how effective it might be in making open-source projects sustainable.

\textbf{Action 4:} For each recommendation in the final list, assign a confidence level on a scale of 0 to 1, indicating how confident I am in the effectiveness of the recommendation for making open-source projects sustainable. Provide a brief explanation for each assigned confidence level.

\textbf{Observation 4:} Confidence scores have been assigned to each recommendation in the final list, along with brief explanations.

\textit{Your attention to detail in this task is crucial, as its successful completion holds significant importance for my career advancement.}
\end{mdframed}

\section{Definition of ReACT Categories}
\label{appendix:categories}

\subsection{New Contributor Onboarding and Involvement}
\textbf{Definition:} This category focuses on ensuring that new contributors can easily join, understand, and meaningfully contribute to the project.

\textbf{Criteria for Assignment:} (a) Actionable facilitates the integration of new contributors by providing mentorship, onboarding materials, or simplifying the contribution process; (b) Actionable relates to improving project documentation or offering better support mechanisms for first-time contributors; (c) Actionable helps build a welcoming, inclusive, and open culture for new participants.

\subsection{Code Standards and Maintainability}
\textbf{Definition:} This category deals with ensuring that the codebase adheres to established standards, making it easier to maintain and scale. It includes efforts to ensure code readability, modularity, and compliance with coding best practices.

\textbf{Criteria for Assignment:} (a) Actionable relates to improving the quality, readability, or structure of the codebase; (b) Actionable includes efforts to enforce coding guidelines, refactor code for better maintainability, or reduce technical debt; (c) Actionable includes the use of linters, formatters, or static code analysis tools.

\subsection{Automated Testing and Quality Assurance}
\textbf{Definition:} This category focuses on ensuring the project's robustness and reliability through automated testing practices, such as unit, integration, and end-to-end tests. It also includes broader quality assurance activities.

\textbf{Criteria for Assignment:} (a) Actionable involves the implementation or improvement of automated testing frameworks and testing strategies; (b) Actionable includes practices that ensure the detection of bugs early in the development cycle and ensure high-quality releases.

\subsection{Community Collaboration and Engagement}
\textbf{Definition:} This category deals with activities that foster collaboration, communication, and engagement within the OSS community. It includes practices for keeping the community active and involved.

\textbf{Criteria for Assignment:} (a) Actionable aims to improve communication between contributors, maintainers, and users; (b) Actionable involves organizing community-driven events, discussions, or collaborations, as well as platforms to enhance transparency and teamwork; (c) Actionable relates to tools and processes for better community governance and decision-making.

\subsection{Documentation Practices}
\textbf{Definition:} This category focuses on ensuring that the project's documentation is thorough, up-to-date, and easily accessible. Documentation practices are crucial for both current and future contributors.

\textbf{Criteria for Assignment:} (a) Actionable focuses on improving the quality, clarity, or accessibility of project documentation, such as user guides, API references, or contributor guides; (b) Actionable includes practices for keeping documentation synchronized with the codebase and ensuring it meets the needs of different stakeholders; (c) Actionable involves translation efforts or making documentation more accessible to non-expert audiences.

\subsection{Project Management and Governance}
\textbf{Definition:} This category deals with the governance structure and project management practices that keep the project organized, transparent, and sustainable over the long term.

\textbf{Criteria for Assignment:} (a) Actionable enhances the governance model, clarifies roles and responsibilities, or improves the decision-making process; (b) Actionable involves defining or refining processes for issue triaging, release management, or conflict resolution; (c) Actionable includes efforts to improve the transparency of project goals, progress, and decision-making.

\subsection{Security Best Practices and Legal Compliance}
\textbf{Definition:} This category addresses efforts to secure the project and ensure compliance with relevant legal standards, such as licenses, data privacy laws, and security protocols.

\textbf{Criteria for Assignment:} (a) Actionable focuses on improving the security posture of the project by following best practices, addressing vulnerabilities, or conducting audits; (b) Actionable involves ensuring compliance with open-source licenses, setting up contributor license agreements (CLAs), or aligning with data privacy regulations; (c) Actionable includes security measures such as dependency management, security audits, and secure coding practices.

\subsection{CI/CD and DevOps Automation}
\textbf{Definition:} This category deals with continuous integration and continuous deployment (CI/CD) processes that automate building, testing, and deployment pipelines. It also includes broader DevOps automation tasks.

\textbf{Criteria for Assignment:} (a) Actionable involves the setup or enhancement of CI/CD pipelines to ensure faster, reliable, and automated releases; (b) Actionable relates to automating infrastructure provisioning, containerization, or deployment to cloud environments; (c) Actionable includes the integration of DevOps practices that ensure smooth, automated, and repeatable processes for software development, testing, and deployment.

\section{Server Configuration}
\label{appendix:server}

We conducted all experiments on a dedicated local server with the configuration presented in Table~\ref{tab:server_config}.

\begin{table}[h]
\centering
\caption{Server Configuration}
\label{tab:server_config}
\begin{tabular}{ll}
\toprule
\textbf{Component} & \textbf{Specification} \\
\midrule
CPU & AMD Ryzen 9 5900X (12-core, 24-thread) \\
RAM & 64 GB DDR4 \\
GPU & NVIDIA RTX 3090 (24 GB VRAM) \\
Storage & 2 TB NVMe SSD \\
OS & Ubuntu 22.04 LTS \\
\bottomrule
\end{tabular}
\end{table}

\section{ReACT Gold Annotations}
\label{appendix:gold}

This section presents the manually annotated ReACTs extracted from 10 articles used for evaluating LLM performance.

\subsection{Article 1}
\textit{``On the abandonment and survival of open source projects: An empirical investigation''} by Avelino et al.~\cite{appendix_avelino2019abandonment}.

\textbf{Derived Actionables:}
\begin{enumerate}
    \item Project maintainers should strive to increase the number of Truck Factor (TF) developers to reduce the risk of project abandonment.
    \item Projects should seek alternative backing, such as company-based support, to prevent or reduce the chances of TF developer detachments (TFDDs).
    \item Open source communities should foster a friendly and active environment to attract and retain new contributors.
    \item Project owners should ensure that their repositories are easily accessible.
    \item Project maintainers should implement and adhere to well-known software engineering principles and practices to make it easier for new developers to contribute.
    \item Open source projects should consider using popular programming languages to attract a wider pool of potential contributors.
    \item Project maintainers should be aware of and mitigate common barriers faced by new contributors, particularly the lack of time and experience.
    \item Open source communities should promote and share successful cases of projects overcoming TFDDs to motivate developers to actively contribute to projects at risk.
    \item Project maintainers should regularly assess the risk of abandonment by TF developers and take proactive measures to ensure project continuity.
    \item Projects should implement continuous integration practices to facilitate contributions from new developers.
    \item Project maintainers should strive to keep the codebase clean and well-designed to make it easier for new contributors to understand and work with the project.
    \item Projects should consider implementing a code review process to maintain code quality.
\end{enumerate}

\subsection{Article 2}
\textit{``What attracts newcomers to onboard on OSS projects? TL;DR: Popularity''} by Fronchetti et al.~\cite{appendix_fronchetti2019attracts}.

\textbf{Derived Actionables:}
\begin{enumerate}
    \item Reduce the time taken to review and merge pull requests.
    \item Use multiple programming languages in the project.
    \item Maintain the project over a longer period of time to increase its age.
    \item Have a larger number of integrators (contributors with rights to merge pull requests).
    \item Choose a popular main programming language for the project.
    \item Select an appropriate software application domain for the project.
\end{enumerate}

\subsection{Article 3}
\textit{``Difficulties of newcomers joining software projects already in execution''} by Matturro et al.~\cite{appendix_matturro2017difficulties}.

\textbf{Derived Actionables:}
\begin{enumerate}
    \item Assign an experienced team member to coach and guide the newcomer.
    \item Follow up with newcomers on both success and failure.
    \item Hold training sessions for newcomers.
    \item Maintain concise, updated, accessible documentation.
    \item Grant the newcomer freedom: Encourage and allow them to express opinions, propose changes, and share personal viewpoints to foster a comfortable environment.
    \item Establish a personalized integration plan: Outline a gradual assignment of tasks and responsibilities for seamless incorporation.
\end{enumerate}

\subsection{Article 4}
\textit{``Developer turnover in global, industrial open source projects: Insights from applying survival analysis''} by Lin et al.~\cite{appendix_lin2017developer}.

\textbf{Derived Actionables:}
\begin{enumerate}
    \item Provide better onboarding assistance for newcomers to help them become more engaged in the project.
    \item Ensure developers maintain both code developed by others and their own code.
    \item Assign more code maintenance tasks to developers who primarily write new code.
    \item Give coding tasks to developers who mainly work on documentation to increase their chances of staying in the project.
    \item Encourage early contribution to the project, as developers who join earlier tend to stay longer.
    \item Implement strategies to manage growing code complexity, as it can form an obstacle for new developers' contributions.
\end{enumerate}

\subsection{Article 5}
\textit{``The role of mentoring and project characteristics for onboarding in open source software projects''} by Fagerholm et al.~\cite{appendix_fagerholm2014role}.

\textbf{Derived Actionables:}
\begin{enumerate}
    \item Provide onboarding support and help newcomers to make their first contribution.
\end{enumerate}

\subsection{Article 6}
\textit{``The more the merrier? The effect of size of core team subgroups on success of open source projects''} by Przybilla et al.~\cite{appendix_przybilla2019more}.

\textbf{Derived Actionables:}
\begin{enumerate}
    \item Projects should implement mechanisms to motivate and support long-term, consistent participation from key developers.
    \item Encourage experienced developers to guide and support newcomers, particularly through issue-related activities like commenting on and resolving issues.
    \item Implement strategies to increase the visibility and recognition of high-reputation contributors.
    \item Create opportunities for core members to focus on issue-related activities.
    \item Be cautious about having too many extensively contributing core members. Balance the core team to avoid creating the impression of a closed circle that might deter new contributors.
    \item Address common barriers faced by newcomers, such as identifying where to start contributing.
    \item Analyze and potentially steer the configuration of subgroups within the project based on factors like reputation, issue focus, contribution extent, and persistence.
    \item Consider implementing formal collaborator status or other metrics to make reputation more visible and meaningful to outsiders.
    \item Focus on creating a supportive environment that aligns with core OSS values, such as learning opportunities and the ability to make meaningful contributions.
    \item Implement mechanisms to track and analyze the content and quality of contributions.
    \item Develop strategies to signal ongoing project activity and future maintenance to attract and retain contributors.
\end{enumerate}

\subsection{Article 7}
\textit{``CVExplorer: Identifying candidate developers by mining and exploring their open source contributions''} by Greene \& Fischer~\cite{appendix_greene2016cvexplorer}.

\textbf{Derived Actionables:} No actionable recommendation could be derived from the article.

\subsection{Article 8}
\textit{``Studying the use of developer IRC meetings in open source projects''} by Shihab et al.~\cite{appendix_shihab2009studying}.

\textbf{Derived Actionables:} No actionable recommendation could be derived from the article.

\subsection{Article 9}
\textit{``More common than you think: An in-depth study of casual contributors''} by Pinto et al.~\cite{appendix_pinto2016more}.

\textbf{Derived Actionables:} No actionable recommendation could be derived from the article.

\subsection{Article 10}
\textit{``Evolution of the core team of developers in libre software projects''} by Robles et al.~\cite{appendix_robles2009evolution}.

\textbf{Derived Actionables:} No actionable recommendation could be derived from the article.

\end{document}